\pdfoutput=1
\documentclass{article}

\usepackage{lineno}
\usepackage{graphicx}
\usepackage{multirow}
\usepackage{subfigure}
\usepackage{atlasphysics}
\usepackage{booktabs} 
\usepackage{epstopdf}
\usepackage[square,comma,sort&compress,numbers]{natbib}
\usepackage{jheppub}
\usepackage{subfigure}

\def\thetitle{Measurement of $ZZ$ production in $pp$ collisions at $\mathbf{\sqrt{s}=7}\TeV$ and limits on anomalous
              $ZZZ$ and $ZZ\gamma$ couplings with the ATLAS detector}
\def\ZZs{$\ZZ^{*}$}
\def\ZZS{$\ZZ^{(*)}$}
\def\missET {$E_{\mathrm{T}}^{\mathrm{miss}}$}

\def\zzllvv {$ZZ\ra\ll\nu\bar{\nu}$}

\def\zzllll {$ZZ\ra\ll\ell'^{+}\ell'^{-}$}
\def\zzsllll {$ZZ^{*}\ra\ll\ell'^{+}\ell'^{-}$}
\def\zzSllll {$ZZ^{(*)}\ra\ll\ell'^{+}\ell'^{-}$}
\def\llvv {$\ll\nu\bar{\nu}$}
\def\llll {$\ll\ell'^{+}\ell'^{-}$}
\def\ztt {$Z\ra\tau^{+}\tau^{-}$}

\def\acer{{\sc AcerMC}}
\def\sherpa {{\sc Sherpa}}
\def\jimmy {{\sc Jimmy}}
\def\ggtwozz {{\sc gg2zz}}
\def\madgraph {{\sc Madgraph}}
\def\powhegbox {{\sc PowhegBox}}

\def\lumi {4.6 \ifb}

\def\fidllll {25.4$^{+3.3}_{-3.0}$ (stat.) $^{+1.2}_{-1.0}$ (syst.) $\pm$ 1.0 (lumi.) \fb}
\def\fidsllll {29.8$^{+3.8}_{-3.5}$ (stat.) $^{+1.7}_{-1.5}$ (syst.) $\pm$ 1.2 (lumi.) \fb}
\def\fidllvv {12.7$^{+3.1}_{-2.9}$ (stat.) $^{+1.7}_{-1.7}$ (syst.) $\pm$ 0.5 (lumi.) \fb}
\def\totzz {6.7 $\pm$ 0.7 (stat.) $^{+0.4}_{-0.3}$ (syst.) $\pm$ 0.3 (lumi.) \pb}
\def\theoryzz {6.18$^{+0.25}_{-0.18}$ \pb}
\def\theoryzzmass {5.89$^{+0.22}_{-0.18}$ \pb}
\def\theoryfidllll {20.9 $\pm$ 0.1 (stat.) $^{+1.1}_{-0.9}$ (theory) \fb}
\def\theoryfidsllll {25.6 $\pm$ 0.1 (stat.) $^{+1.3}_{-1.1}$ (theory) \fb}
\def\theoryfidllvv {12.5 $\pm$ 0.1 (stat.) $^{+1.0}_{-1.1}$ (theory) \fb}

\usepackage{preprintcover}  
\PreprintCoverPaperTitle{Measurement of $ZZ$ production in $pp$ collisions at $\mathbf{\sqrt{s}=7}\TeV$ and limits on anomalous
              $ZZZ$ and $ZZ\gamma$ couplings with the ATLAS detector}
\PreprintIdNumber{CERN-PH-EP-2012-318}
\PreprintCoverAbstract{A measurement of the \ZZ production cross section in proton--proton collisions
at $\sqrt{s} = 7\TeV$ using data recorded by the ATLAS experiment
at the Large Hadron Collider is presented.
In a data sample corresponding to an integrated luminosity of
\lumi\ collected in 2011, events are selected that are consistent either with two $Z$ bosons
decaying to electrons or muons or with one $Z$ boson
decaying to electrons or muons and a second $Z$ boson decaying to
neutrinos. The \zzSllll\ and \zzllvv\ cross
sections are measured in restricted phase-space regions.
These results are then used to derive the total cross section for \ZZ\ events
produced with both \Z\ bosons in the mass range 66 to 116 \GeV, $\sigma_{ZZ}^\mathrm{tot} = $ \totzz,
which is consistent with the Standard Model
prediction of \theoryzzmass\ calculated at next-to-leading order in QCD.
The normalized differential cross sections in bins of various kinematic variables are presented.
Finally, the differential event yield as a function of the transverse momentum of the leading $Z$ boson
is used to set limits on anomalous neutral triple gauge boson couplings in $ZZ$ production.
}
\PreprintJournalName{JHEP}  

\title{\thetitle}

\begin{document}

\date{\today}

\abstract{
A measurement of the \ZZ production cross section in proton--proton collisions
at $\sqrt{s} = 7\TeV$ using data recorded by the ATLAS experiment 
at the Large Hadron Collider is presented.  
In a data sample corresponding to an integrated luminosity of
\lumi\ collected in 2011, events are selected that are consistent either with two $Z$ bosons
decaying to electrons or muons or with one $Z$ boson
decaying to electrons or muons and a second $Z$ boson decaying to
neutrinos. The \zzSllll\ and \zzllvv\ cross 
sections are measured in restricted phase-space regions. 
These results are then used to derive the total cross section for \ZZ\ events 
produced with both \Z\ bosons in the mass range 66 to 116 \GeV, $\sigma_{ZZ}^\mathrm{tot} = $ \totzz, 
which is consistent with the Standard Model
prediction of \theoryzzmass\ calculated at next-to-leading order in QCD.
The normalized differential cross sections in bins of various kinematic variables are presented.
Finally, the differential event yield as a function of the transverse momentum of the leading $Z$ boson 
is used to set limits on anomalous neutral triple gauge boson couplings in $ZZ$ production.
}

\maketitle

\section{Introduction}\label{sec:Introduction}
        The production of pairs of \Z\ bosons at the Large Hadron Collider (LHC) provides an excellent
opportunity to test the predictions of the electroweak sector of the Standard Model (SM)
at the TeV energy scale.
In the SM, $Z$ boson pairs can be produced via non-resonant processes or in the decay of Higgs bosons. 
Deviations from SM expectations for the total or differential \ZZ\ production cross sections 
could be indicative of the production of new resonances decaying to \Z\ bosons or other non-SM contributions. 

\begin{figure}[htbp]
  \begin{center}
  \subfigure[]{
  \includegraphics[width=0.30\textwidth]{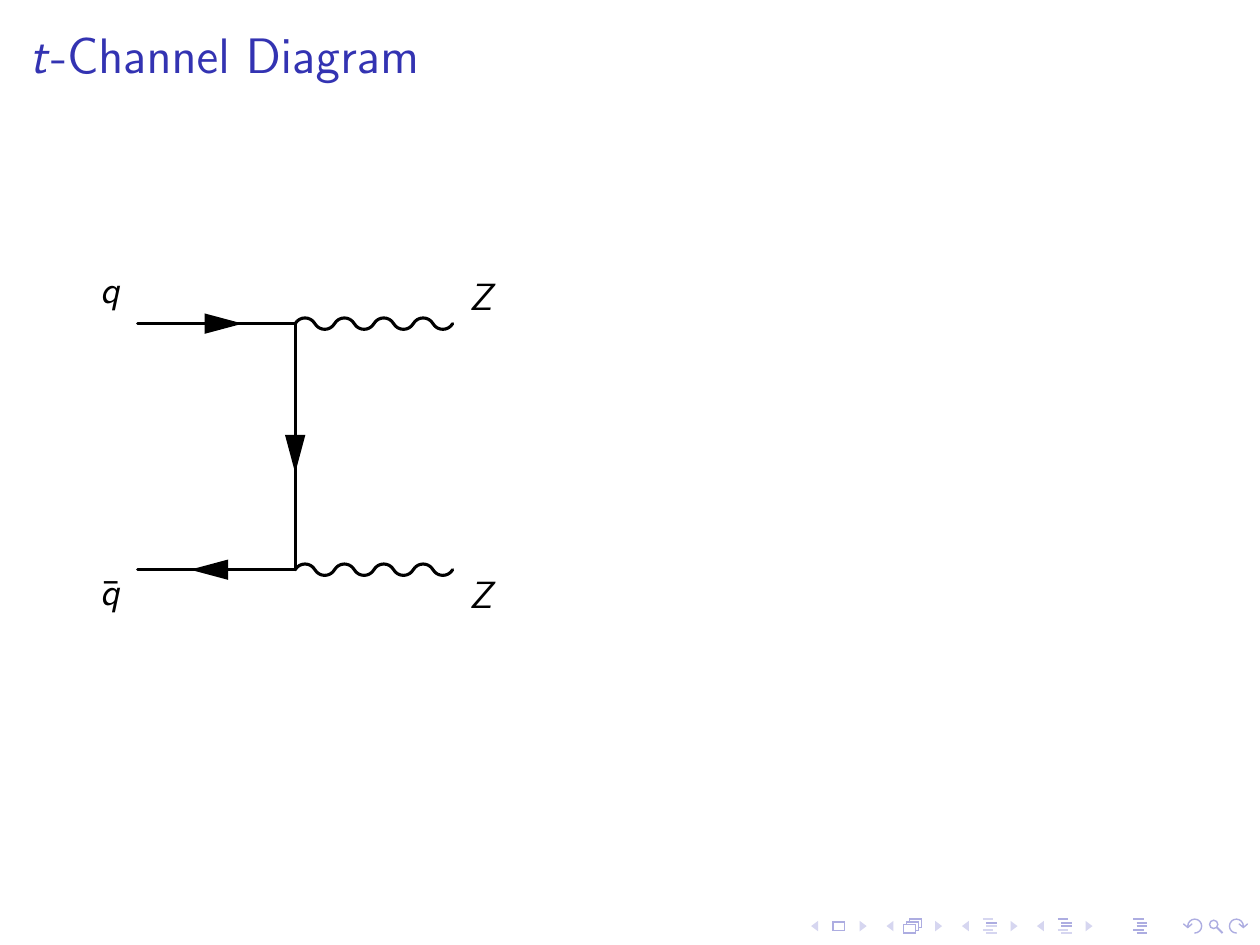}
  }
  \subfigure[]{
  \includegraphics[width=0.30\textwidth]{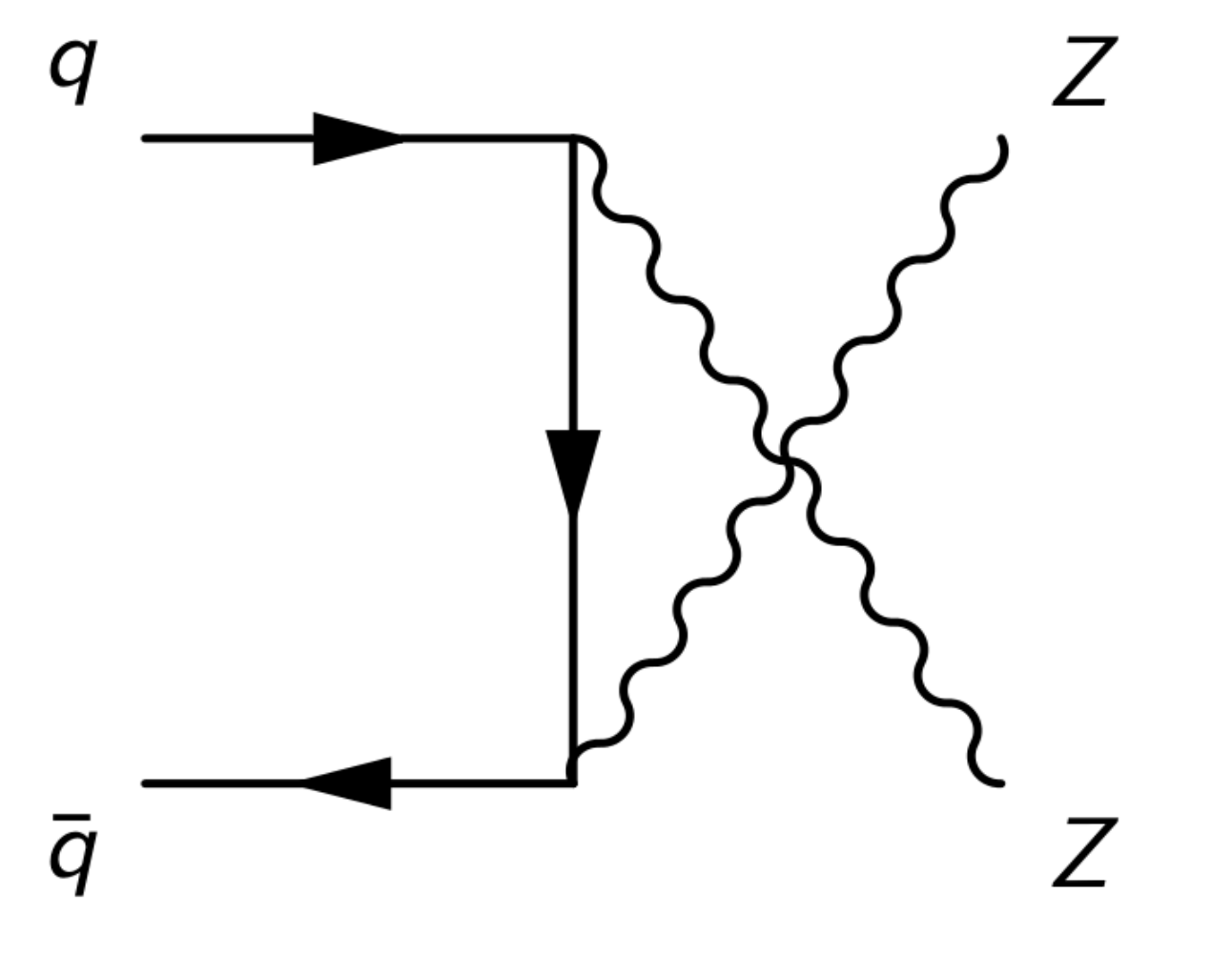}
  }
  \subfigure[]{
  \includegraphics[width=0.30\textwidth]{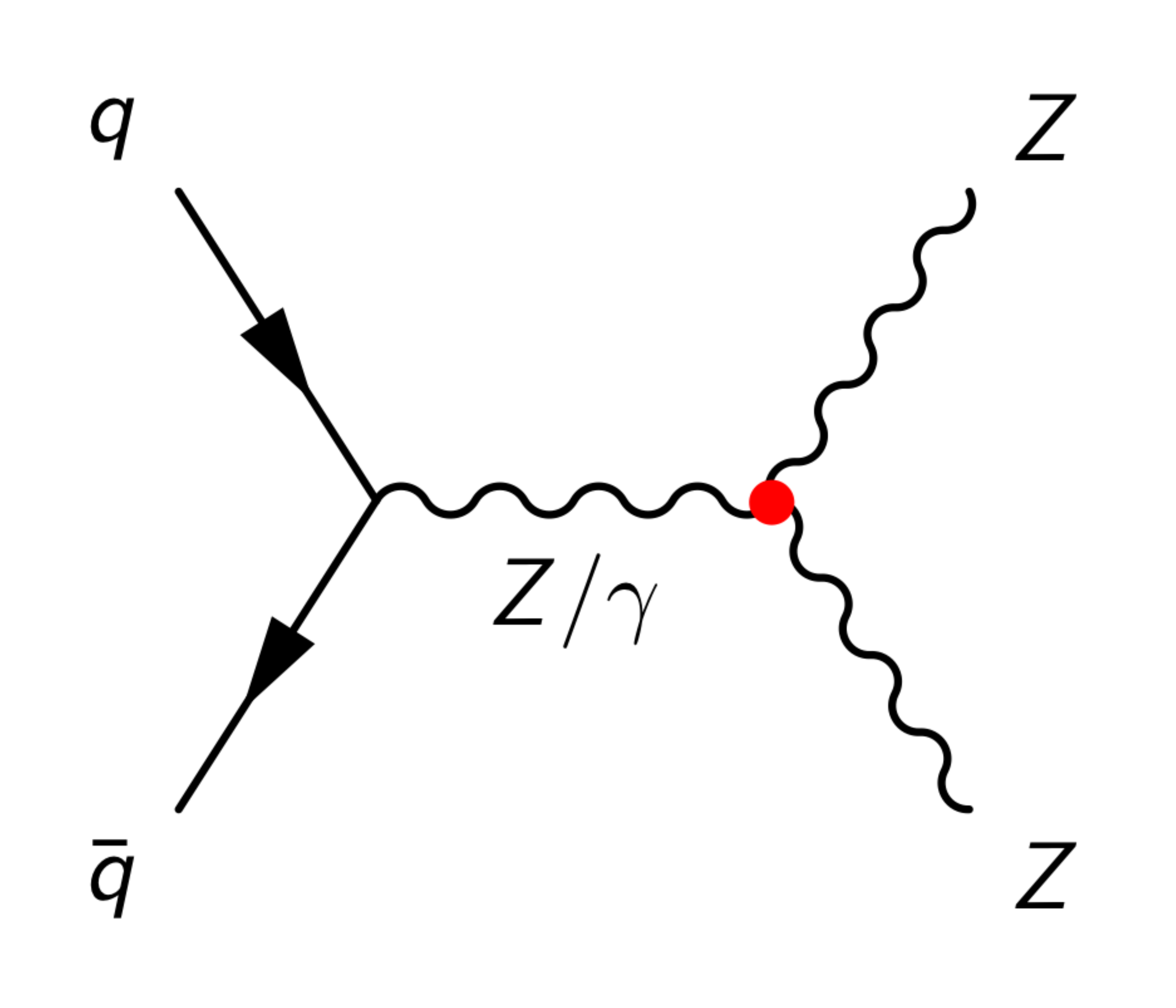}\hfill
  }
  \subfigure[]{
  \includegraphics[width=0.30\textwidth]{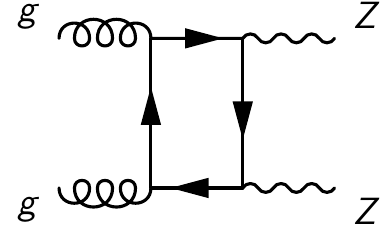}
  }
  \subfigure[]{
  \includegraphics[width=0.30\textwidth]{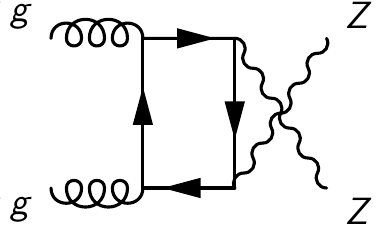}\hfill
  }
  \caption{\label{fig:feyndiag}Leading order Feynman diagrams for \ZZ production through the $q\bar{q}$ and $gg$ initial state at hadron 
           colliders. The $s$-channel diagram, (c), contains the $ZZZ$ and $ZZ\gamma$ neutral TGC vertices which do not exist in the SM.} 
  \end{center}
\end{figure}

Non-resonant \ZZ\ production proceeds at leading order (LO) via $t$- and $u$-channel quark--antiquark  
interactions, while about $6\%$ of the production proceeds via gluon fusion.
The $ZZZ$ and $ZZ\gamma$ neutral triple gauge boson couplings (nTGCs) are absent in the SM, 
hence there is no contribution from $s$-channel \qqbar\ annihilation at tree level. These different 
production processes are shown in figure~\ref{fig:feyndiag}. 
At the one-loop level, nTGCs generated by fermion triangles have a magnitude of the order of $10^{-4}$~\cite{Gounaris:2000tb}. 
Many models of physics beyond the Standard Model
predict values of nTGCs at the level of $10^{-4}$ to $10^{-3}$~\cite{Ellison:1998}. 
The primary signatures of non-zero nTGCs are an increase in the \ZZ cross section at high \ZZ invariant mass 
and high transverse momentum of the \Z\ bosons~\cite{Baur:2000ae}. 
\ZZ production has been studied in \epem\ collisions at 
LEP~\cite{Barate:1999jj,Abdallah:2003dv,Acciarri:1999ug,Abbiendi:2003va,bib:LEPEW2006}, 
in $p\overline{p}$ collisions at the Tevatron~\cite{bib:D0_ZZ1,bib:CDF_ZZ,bib:D0_ZZ2,bib:D0_ZZ3}
 and recently in $pp$ collisions at the LHC~\cite{ATLAS_ZZ4l:1fb2011,ZZPaperCMS}.
No deviation of the measured total cross section from the SM expectation has been observed, 
and limits on anomalous nTGCs have been set~\cite{bib:LEPEW2006,bib:D0_ZZ1,ATLAS_ZZ4l:1fb2011,ZZPaperCMS}. 
In searching for the SM Higgs boson, the ATLAS and CMS collaborations observed recently a neutral boson resonance with a mass around 126~GeV~\cite{ATLAS_Higgs:2012gk, CMS_Higgs:2012gu, bib:higgs7tev}. 
A SM Higgs boson with that mass 
can decay to two $Z$ bosons only when at least one $Z$ boson is off-shell, and even in this case, the contribution is less than 3\%.
Searches for high-mass non-SM \ZZ\ resonances have not resulted in any excess above the SM expectations~\cite{ATLAS:2012iua}.

This paper presents a measurement of 
\ZZ\ production\footnote{Throughout this paper \Z\ should be taken            
to mean $\Z/\gamma^{*}$ when referring to decays to charged leptons, and just \Z\ when referring to decays to neutrinos. 
} 
in proton--proton collisions at a centre-of-mass energy $\sqrt{s}=7$ \TeV\ using  
\lumi\ 
of integrated luminosity collected by the ATLAS detector at
the LHC.
\ZZ\ events are selected in two channels\footnote{$\ell$ 
represents either electrons or muons. $\ell$ and $\ell'$ are used to denote
leptons from a different \Z\ parent, but not necessarily of different flavour.
Decay modes mentioned with the use of $\ell$ indicate the sum of the decay modes
with specific lepton flavours.}: \llll\ and \llvv. Two selections are used
in the four-charged-lepton channel: an on-shell \ZZ\ selection denoted by
\zzllll\ where both $Z$ bosons are required to be within the mass range 66--116 
GeV\footnote{Throughout this paper, the 66--116 GeV mass range is referred to as the \Z\ mass window.} 
 and a selection which includes an off-shell $Z$ boson denoted by \zzsllll\ where one $Z$ boson is 
required to be within this mass range and the other can be off-shell and have any mass above 20 GeV. 
In the \llvv\ channel, the $\nu\bar{\nu}$ system is expected to be produced by an off-shell \Z\ boson 
in 2.6\% of the events.
Since this fraction is small and only one event selection is used for this channel, it is referred 
to as \zzllvv\ throughout the paper. 
The \zzSllll\ channel has an excellent signal-to-background ratio,  
but it has a branching fraction six times lower than the \zzllvv\ channel; the
latter has higher background contributions with an expected signal-to-background
ratio around one (after applying the event selections described below).
This paper presents the total \ZZ\ production cross section, 
the fiducial cross section in a restricted phase space for each decay channel (integrated, and as a
function of kinematic parameters for the \ZZ\ selections) and limits on anomalous \mbox{nTGCs} using the observed \ZZ\ event yields as a function of the transverse momentum 
of the leading $Z$ boson\footnote{Leading $Z$ refers to the $Z$ with the higher transverse momentum in \zzllll\ decays or to the
$Z$ boson decaying to a charged lepton pair in \zzllvv\ decays.}.
The results presented in this paper supersede the previously published results~\cite{ATLAS_ZZ4l:1fb2011} which were derived with the first 1.02 \ifb\ of the dataset used here, only with the \zzllll\ decay channel and with the use of the total \ZZ\ event count for the derivation of the limits on anomalous \mbox{nTGCs}.

The total cross section for non-resonant \ZZ\ production is
predicted at next-to-leading order (NLO) in QCD to be \theoryzz, where the quoted theoretical uncertainties result 
from varying the factorization and renormalization scales simultaneously by a factor of two whilst using the 
full CT10 parton distribution function (PDF) error set~\cite{bib:ct10}. The cross section is calculated 
in the on-shell (zero-width) approximation using MCFM~\cite{Campbell:2011} with CT10;  
it includes a $5.8\%$ contribution from gluon fusion.
When the natural width of the $Z$ boson is used and both $Z$ bosons are required to be within the \Z\ mass window, 
the NLO cross section is predicted to be \theoryzzmass. 
The cross sections given here are calculated at a renormalization and factorization scale equal to half the mass 
of the diboson system. 
The total cross section using the zero-width approximation was previously measured to be 
8.5$^{+2.7}_{-2.3}$ (stat.) $^{+0.4}_{-0.3}$ (syst.) $\pm$ 0.3 (lumi.) \pb~\cite{ATLAS_ZZ4l:1fb2011}.

This paper is organized as follows: 
an overview of the ATLAS detector, data, signal and background Monte Carlo (MC)
samples used for this analysis is given in section~\ref{sec:ATLAS}; 
section~\ref{sec:Reconstruction} describes the selection of the physics objects; 
section~\ref{sec:SignalAcceptance} describes the fiducial phase space of the 
measurement, the corresponding \ZZ\ cross section definition and the acceptances
of the event selection and fiducial phase space;
section~\ref{sec:Background} explains how the backgrounds to the
\llll\ and \llvv\ final states are estimated with a combination of simulation and data-driven techniques;
section~\ref{sec:Results} presents the results: cross section, differential cross sections and nTGC limits; finally, a summary of the main results is given in section \ref{sec:Conclusions}.

\section{The ATLAS detector and data sample}\label{sec:ATLAS}
        The ATLAS detector~\cite{bib:ATLASDetectorPaper} is a multipurpose particle detector with a cylindrical geometry. It consists of inner tracking devices 
surrounded by a superconducting solenoid, electromagnetic and
hadronic calorimeters and a muon spectrometer with a toroidal magnetic field. The inner
detector, in combination with the 2\,T field from the solenoid, provides precision tracking of charged particles in the pseudorapidity 
range $|\eta|<2.5$\footnote{ATLAS uses a right-handed coordinate
system with its origin at the nominal interaction point in the centre of the detector and
the $z$-axis along the beam direction. The $x$-axis points from the interaction point to
the centre of the LHC ring, and the $y$-axis points upwards. Cylindrical coordinates ($r$,$\phi$)
are used in the transverse plane, $\phi$ being the azimuthal angle around the beam direction.
The pseudorapidity $\eta$ is defined in terms of the polar angle $\theta$ as
$\eta = - \ln\tan(\theta/2)$.}.
It consists of a
silicon pixel detector, a silicon microstrip detector and a straw tube tracker that also provides 
transition radiation measurements for electron identification in the pseudorapidity range $|\eta| < 2.0$. 
The calorimeter system  covers the pseudorapidity range $|\eta| < 4.9$. 
The electromagnetic calorimeter uses liquid argon (LAr) as the active material with lead as an absorber ($|\eta|<3.2$). 
It identifies electromagnetic showers and measures their energy and position;
in the region $|\eta|<2.5$ it is finely segmented and provides electron identification in conjunction with the inner detector
which covers the same $\eta$ region. 
Hadronic showers are measured in the central rapidity range ($|\eta|<1.7$) by scintillator tiles with iron absorber, 
while in the end-cap region ($1.5<|\eta|<3.2$) a LAr calorimeter with a copper absorber is used.
In the forward region ($3.2<|\eta|<4.9$) a LAr calorimeter with a copper absorber for the first layer and tungsten for 
the last two layers is used for both electromagnetic and hadronic showers.
All calorimeters are used to measure jets. 
The muon spectrometer surrounds the calorimeters; it consists of superconducting air-core toroid magnets, high-precision tracking chambers which provide muon identification and tracking measurement in the pseudorapidity range $|\eta|<2.7$, and separate trigger chambers covering $|\eta|<2.4$.

A three-level trigger system selects events to be recorded for offline analysis. 
The events used in this analysis were selected with  single-lepton triggers with nominal transverse
momentum (\pT) thresholds of 20 or 22 \GeV\ (depending on the instantaneous luminosity of the LHC) for electrons and 18 \GeV\ for muons. 
The efficiencies of the single-lepton triggers have been determined as a function of lepton pseudorapidity and transverse momentum using 
large samples of $\Z\ra\ll$ events. 
The trigger efficiencies for events passing the offline selection described below are all greater than 98\%.

The measurements presented here uses the full data sample of proton--proton collisions at $\sqrt{s} = 7 \TeV$
recorded in 2011. 
After data quality
requirements, the total integrated luminosity used in the analysis is \lumi\ with an 
uncertainty of 3.9\%~\cite{bib:ATLASLumi2010}.

        \subsection{Simulated data samples}\label{sec:MCsample}
        Monte Carlo simulated samples cross-checked with data are used to calculate several quantities 
used in this measurement, including acceptance, efficiency and some of the
background to the \zzllvv\ 
decay channel. 
The NLO generator \powhegbox~\cite{bib:powhegbox,bib:powhegboxzz} with the CT10 
 PDF set, interfaced to \pythia~\cite{bib:pythia}, is used to model 
the signal for both channels. 
The LO multi-leg generator \sherpa~\cite{bib:sherpa} with the CTEQ6L1
 PDF set~\cite{bib:cteq6li} in comparison with \powhegbox\ is used to 
evaluate systematic uncertainties. The
contribution from $gg\ra ZZ$ is modelled by the
 \ggtwozz\ generator~\cite{Binoth:2008pr} interfaced to
 \herwig~\cite{bib:Herwig} to model parton showers and to \jimmy~\cite{bib:jimmy}  for multiparton interactions.
In each case, 
the simulation includes the interference terms between the \Z\ and $\gamma^{*}$ diagrams. 
For both the \llll\ and \llvv\ final states, MCFM is used to calculate theoretical uncertainties, and 
 \sherpa\ is used for the generation of signal samples with neutral triple gauge couplings. 

The LO generator \alpgen~\cite{bib:alpgen} with CTEQ6L1 PDFs is used to simulate $Z$+jets, 
$W$+jets, $Z\gamma$ and $W\gamma$ background 
events with \jimmy\ used for multiparton interactions and \herwig\ for parton showers. 
The NLO generator \mcatnlo~\cite{bib:mcatnlo} with CT10 PDFs is used to model
$t\bar{t}$ background processes as well as $WW$ production. 
The single-top $Wt$ process is modelled with \acer~\cite{bib:acer} with the MSTW2008 PDFs~\cite{bib:MSTW2008}.
The LO generator \herwig\ with MSTW2008 PDFs is used to
model $WZ$ production. The LO generator \madgraph~\cite{bib:madgraph} with CTEQ6L1 PDFs is also 
used to model $Z\gamma$ and $W\gamma^{*}$ events, where \pythia\ is used for hadronization and
showering.

The detector response is simulated~\cite{bib:ATLASMCPaper} with a program based on {\sc Geant4}~\cite{bib:geant4a}.
Additional inelastic $pp$ events are
included in the simulation, distributed so as to reproduce the number of collisions per 
bunch-crossing in the data. The detector response to interactions in the out-of-time bunches from pile-up is also modelled in the simulation.
The results of the simulation are corrected with scale factors determined by
comparing efficiencies observed in data to those in the
simulated events, and the lepton momentum scale and resolution
are finely adjusted to match the observed dilepton spectra in $Z\rightarrow\ell\ell$ events using a sample of $Z$ bosons.

\section{Event reconstruction and selection}\label{sec:Reconstruction}
        Events are required to contain a primary vertex
formed from at least three associated tracks with \mbox{$\pT>400\MeV$}.

\subsection{Leptons, jets and missing energy}

\subsubsection{Common lepton selection}

Muons are identified by matching tracks (or track segments) reconstructed in the muon spectrometer 
to tracks reconstructed in the inner detector~\cite{bib:ATLAS_W_Z}. 
The momenta of these combined muons are calculated by combining the information from the two systems
and correcting for the energy deposited in the calorimeters. 
The analyses of both decay channels use muons which have full tracks reconstructed in 
the muon spectrometer with $\pT >20\GeV$ and $|\eta| < 2.5$. 
The \zzSllll\ channel  recovers additional $ZZ$ acceptance with
minimal additional background using a lower
threshold of $\pT > 7\GeV$ 
and by accepting muons with segments reconstructed in the muon spectrometer (in this latter case, 
the muon spectrometer is used to identify the track as a muon, but its momentum is measured using the inner detector; for the purposes of the discussion 
below, these muons are also referred to as {\emph {combined}} muons). 

Electrons are reconstructed from an energy cluster in the electromagnetic calorimeter matched to a track in the 
inner detector~\cite{bib:ATLAS_W_Z}; the transverse momentum is computed from the calorimeter energy and the direction from the track parameters measured in the inner detector.  
The electron track parameters are corrected for
bremsstrahlung energy loss using the Gaussian-sum filter algorithm~\cite{GSF}. 
 Electron candidates in the \zzSllll\ ($ZZ\to \ll \nu\bar{\nu}$) channel are required to 
 have longitudinal and transverse shower profiles consistent with those expected from
 electromagnetic showers, by satisfying the 
loose (medium)
identification criteria described in ref.~\cite{bib:egammaperf} reoptimized for the 2011 data-taking conditions.
They are also required to have a transverse momentum 
of at least 7 (20) \GeV\ and a pseudorapidity of $|\eta| < 2.47$. 

In order to reject non-prompt leptons from the decay of heavy quarks and 
fake electrons from misidentified jets (charged hadrons or photon conversions), all selected leptons 
must satisfy isolation requirements based on calorimetric and tracking information and must be 
consistent with originating from the primary vertex.
For the calorimetric isolation the scalar sum of the transverse energies, $\Sigma \ET$, of calorimeter
deposits inside a cone around the lepton, 
corrected to remove the energy from the lepton and from additional interactions (pile-up), is formed.
In the \zzSllll\ ($ZZ\to \ll \nu\bar{\nu}$) channel,
the $\Sigma \ET$ inside a cone of size $\Delta{R}=\sqrt{\left(\Delta\phi\right)^2 + \left(\Delta\eta\right)^2} = $ 0.2 (0.3) around the 
lepton is required to be no more than 30\% (15\%) of the lepton \pT.
For the track isolation, the scalar sum of the transverse momenta, $\Sigma \pT$, of
inner detector tracks inside a cone of size $\Delta{R} =$ 0.2 (0.3) 
around the lepton is required to be no more than 15\% of the lepton \pT.
The wider cone size, in conjunction with the same or tighter requirements on the fraction of extra activity allowed in the cone, 
corresponds to more stringent isolation requirements applied to the \zzllvv\ channel compared to the \zzSllll\ channel. 
This reflects the need to reduce the much higher reducible background (predominantly from $Z+$jets, $t\bar{t}$ and $WW$).
To ensure that the lepton originates from the primary vertex, its longitudinal impact parameter $|z_{0}|$ is required to be less than 2 mm, and its transverse impact parameter significance (the transverse impact parameter divided by its error), $|d_{0}/\sigma_{d_{0}}|$, 
is required to be less than 3.5 (6) for muons (electrons).
Electrons have a worse impact parameter resolution than muons due to bremsstrahlung.

Since muons can radiate photons which may then convert to electron-positron pairs,
electron candidates within $\Delta R = 0.1$ of any selected muon
are not considered. If two electron candidates are within $\Delta R = 0.1$ of each other, the one with
the lower \pT\ is removed.

\subsubsection{Extended-lepton selection}

Two additional categories of muons are considered for the \zzSllll\ channel:
forward spectrometer muons with $2.5 < |\eta| < 2.7$ (in a region outside the nominal coverage of the inner detector)
and calorimeter-tagged muons with $|\eta| < 0.1$ (where there is a limited geometric coverage in the
muon spectrometer).
Forward spectrometer muons are required to have a full track that is reconstructed in the muon spectrometer;
if these muons are also measured in the inner detector, their momentum is measured using the combined information; otherwise, only the muon spectrometer information is used.
In either case, such muons are required to have $\pT > 10\GeV$ and the
$\Sigma \ET$ of calorimeter deposits inside a cone of
size $\Delta R = 0.2$ around the muon is required to be no more than 15\% of the
muon \pT, while no requirement is made on $\Sigma \pT$.
The same impact parameter requirements as for the combined muons are imposed for the forward muons measured 
in the inner detector; no such requirement is imposed on those measured in the muon spectrometer only. 
Calorimeter-tagged muons are reconstructed from calorimeter energy deposits consistent with a muon which are matched to an inner detector track with $\pT > 20\GeV$
and are required to satisfy the same impact parameter and isolation criteria as for the combined muons.

The \zzSllll\ channel also uses calorimeter-only
electrons with $2.5 < |\eta| < 3.16$ and $\pT > 20\GeV$ passing the tight identification requirements~\cite{bib:egammaperf} for this forward $\eta$ region, where only the longitudinal and transverse shower profiles in the calorimeters are used for their identification.
Their transverse momentum is computed from the calorimeter energy and the electron direction, where the electron direction is computed using the primary vertex position and the shower barycentre position in the calorimeter.
Being identified outside the acceptance of the inner detector, no impact parameter requirements can be applied 
to these calorimeter-only electron candidates, and their charge is not measured. 
Since only one such electron is allowed in the event, and since all other leptons have their charge measured, the 
calorimeter-only electron is assigned the charge needed to have two pairs of same-flavour opposite-sign leptons in the event.
The requirements described above constrain the additional background introduced by the inclusion of calorimeter-only electrons, and no isolation requirements are imposed on such electrons.

The use of the extended-lepton selection
increases the  \zzllll\ and \zzsllll\ acceptance
by about 6\% from the forward spectrometer muons,  4\% from the calorimeter-tagged muons and 6\% from the forward electrons.
The expected background is kept small by requiring
each event to have at most one lepton from each extended-lepton category, and each such lepton to be paired with a non-extended lepton.

\subsubsection{Jets and missing transverse momentum}

For the \zzllvv\ selection, events which contain 
at least one well-reconstructed jet are vetoed to reduce background from top-quark production.
Jets are reconstructed from topological clusters of energy in the calorimeter~\cite{bib:topo} using the anti-$k_{t}$ 
algorithm~\cite{bib:antikt4} with radius parameter $R=0.4$. 
The measured jet energy is corrected for detector inhomogeneities and for the non-compensating
nature of the calorimeter using \pT- and $\eta$-dependent correction factors based on
Monte Carlo simulations with adjustments from in-situ measurements~\cite{bib:jetcalibPhoton,bib:jetcalibZ}. 
Jets are required to have $\pT > 25 \GeV$ and $|\eta| < 4.5$.
In order to minimize the impact of jets from pile-up at high luminosity, the jet vertex fraction is required to be at 
least 0.75; the jet vertex fraction is defined as the sum 
of the \pT\ of tracks associated to the jet and originating from the primary vertex, 
divided by the sum of the \pT\ of all the tracks associated to the jet. 
If a reconstructed jet and a lepton lie within $\Delta{R}=0.3$ of each other, the jet is not considered in the analysis.

The missing transverse momentum \missET\ is the imbalance of transverse momentum in the event. A large imbalance in the transverse 
momentum is a signature of the \zzllvv\ decay channel. The two-dimensional \missET\ vector 
is determined from the negative vectorial sum of reconstructed electron, muon and jet momenta together with calorimeter cells not 
associated to any object~\cite{metwznote}. Calorimeter cells are 
calibrated to the jet energy scale if they are associated with a jet and to the electromagnetic energy scale otherwise.
Using calorimeter timing and shower shape information, events that contain jets with $\pT>20$ GeV and not originating 
from proton-proton collisions but from e.g. calorimeter signals due to noisy cells are rejected. 

\subsection{\zzSllll\ selection}

\zzSllll\ events are characterized by four high-\pT, isolated electrons or muons, 
in three channels: \ee\ee, \mumu\mumu\ and \ee\mumu.
Selected events are required to have exactly four leptons and to have passed at least a  
single-muon or single-electron trigger. 
Each combination of lepton pairs is required to satisfy
$\Delta{R}(\ell_1,\ell_2)>0.2$, where $\ell_1$ and $\ell_2$ are used hereafter
to denote a pair of distinct leptons, independent of their \Z\ parent assignment, flavour and charge.
To ensure high and well-measured trigger efficiency, at least one lepton must have $\pT > 20 \GeV\ (25 \GeV)$ for 
the offline muon (electron) and be matched to a muon (electron) reconstructed online by the trigger system within $\Delta R =$ 0.1 (0.15).

Same-flavour, oppositely-charged lepton pairs are combined to form \Z\ candidates.
An event must contain two such pairs. 
In the \ee\ee\ and \mumu\mumu\ channels, ambiguities are resolved by choosing the combination which
results in the smaller value of the sum of $|m_{\ll} - \mZ|$ for the two pairs, where $m_{\ll}$ 
is the mass of the dilepton system and \mZ\ is the mass of the $Z$ boson~\cite{PDG}.
Figure~\ref{fig:mass2d} shows the correlation between the invariant mass 
of the leading (higher \pT) and the sub-leading (lower \pT) lepton pair.
The events cluster in the region where both masses are around \mZ.
At least one lepton pair is required to have invariant mass within the \Z\ mass window, $66 < m_{\ell^{+}\ell^{-}} < 116$ GeV.
If the second lepton pair satisfies
this as well, the event is classified as a $ZZ$ event; if the second
pair satisfies $m_{\ll} > 20$ GeV, the event is classified as a
$ZZ^*$  event. 

With the selection described here, 84 \zzsllll\ candidates are observed, out of
which 66 are classified as \zzllll\ candidates. From the 84 (66) \zzsllll\
(\zzllll) candidates, 8 (7) candidates contain extended leptons.

 \begin{figure}[htbp]
 \begin{center}
  \includegraphics[width=0.7\textwidth]{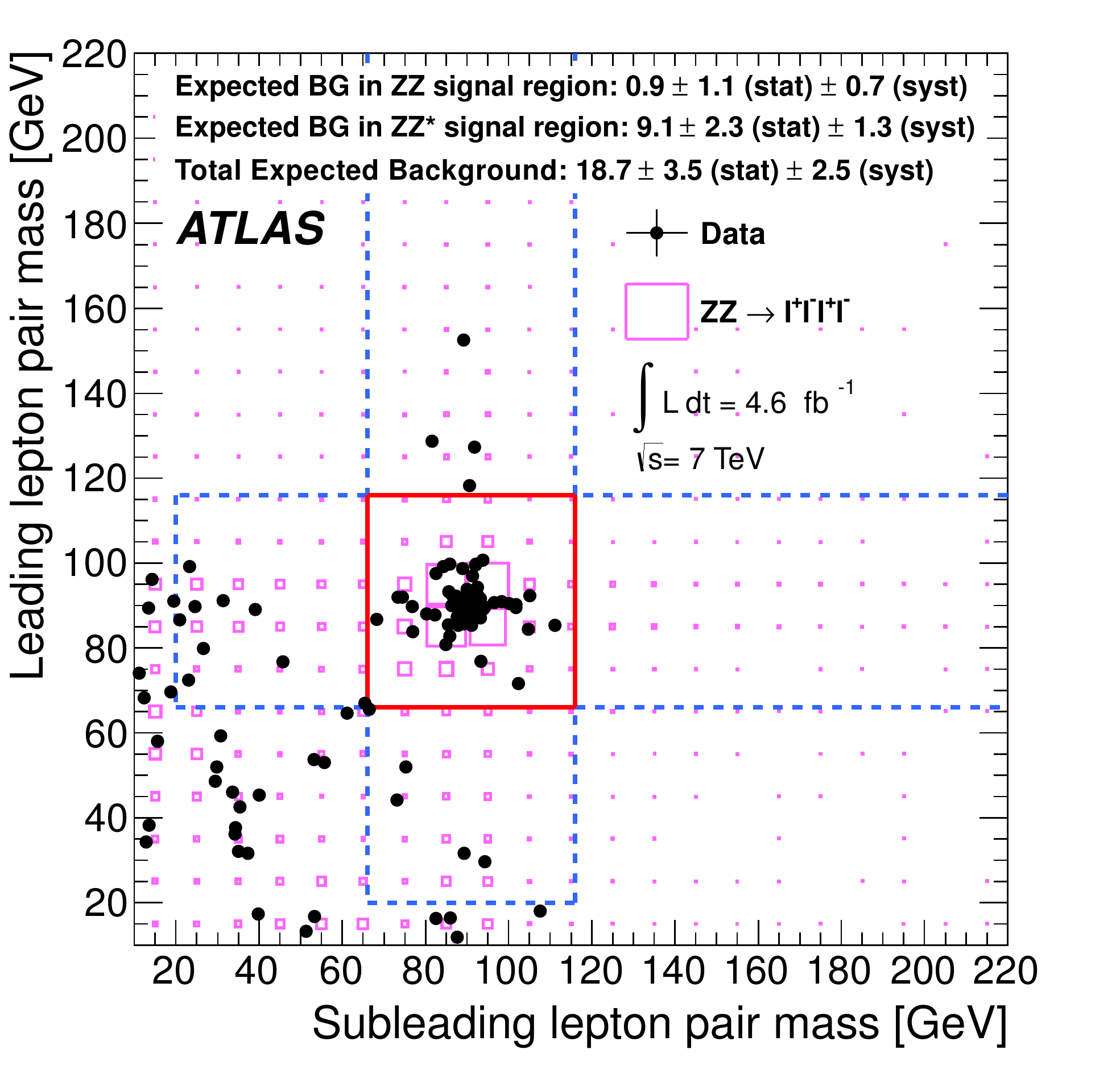}\hfill
  \caption{\small The mass of the leading lepton pair versus the mass of the sub-leading lepton pair.
  The events observed in the data are shown as solid circles and the \zzSllll\ signal prediction from simulation as boxes.
  The size of each box is proportional to the number of events in each bin.  
  The region enclosed by the solid (dashed) lines indicates the signal region defined by the
  requirements on the lepton-pair masses for $ZZ$ ($ZZ^*$) events, as
  defined in the text. 
   }
 \label{fig:mass2d}
 \end{center}
 \end{figure}

\subsection{$ZZ \to \ll \nu\bar{\nu}$ selection}

$ZZ\to \ll \nu\bar{\nu}$  events are characterized by large missing
transverse momentum and two high-\pT, isolated electrons or muons. 
Selected events are required to have exactly two leptons of the same
flavour with $76 < m_{\ll} < 106\GeV$ and to have passed at least a  
single-muon or a single-electron trigger. The mass window is chosen to be 
tighter than the mass window used for the \zzSllll\ channel in order to 
reduce the background from $t\bar{t}$ and $WW$. 
The lepton pair is required to have $\Delta{R}(\ell^{+},\ell^{-})>0.3$. This requirement reflects the choice of the isolation cone for the leptons. 
The same trigger matching requirement as in the \zzSllll\ channel is used.

The \zzllvv\ decay channel analysis makes use of several selections to reduce background. 
The largest background after the mass window requirement consists of $Z+$jets events, which are associated with non-zero 
missing transverse momentum when the \missET\ is mismeasured or when a $b$-quark decays to leptons and 
neutrinos inside of a jet. Since the $Z$ bosons tend to be produced back-to-back, the 
axial-\missET\ (defined as the projection of the \missET\ along the direction opposite to the 
$Z\to\ll$ candidate in the transverse plane) is a powerful 
variable to distinguish \zzllvv\ decays from $Z+$jets. The axial-\missET\ is given by 
$-\vec{E}_{\mathrm{T}}^{\mathrm{miss}} \cdot \vec{p}^Z / p_{\mathrm T}^Z$, where $p_{\mathrm T}^{Z}$ is 
the magnitude of the transverse momentum of the $Z$ candidate. Similarly, the fractional \pT\ difference, 
$ | E_{\mathrm{T}}^{\mathrm{miss}} -p_{\mathrm T}^Z|/p_{\mathrm T}^Z$ is a good variable to 
distinguish the two. 
The axial-\missET\ and fractional \pT\ difference are shown in figure~\ref{fig:nminusone_llvv}. 
In order to reduce $Z+$jets background, the axial-\missET\ must be greater than 75 GeV, and 
the fractional \pT\ difference must be less than 0.4.
To reduce background from top-quark production, events which contain at least one reconstructed jet with 
$p_{\mathrm T}>25$ GeV and $|\eta|<4.5$ are rejected. 

To reduce background from $WZ$ production, 
events with a third lepton (electron or muon) 
with $p_{\mathrm T}$ greater than 10 GeV are rejected. 
The shape of the jet multiplicity distribution is well 
modelled in Monte Carlo simulation as shown in figure~\ref{fig:kindists_jet} for the \zzllll\ and \zzllvv\ 
selections, however, there is an overall excess of about 20\% in the \zzllll\ selection. 
With this selection, 87 \zzllvv\ candidates are observed in data.

\begin{figure}[htbp]
\centering
\subfigure[]{
\includegraphics[width=0.47\textwidth]{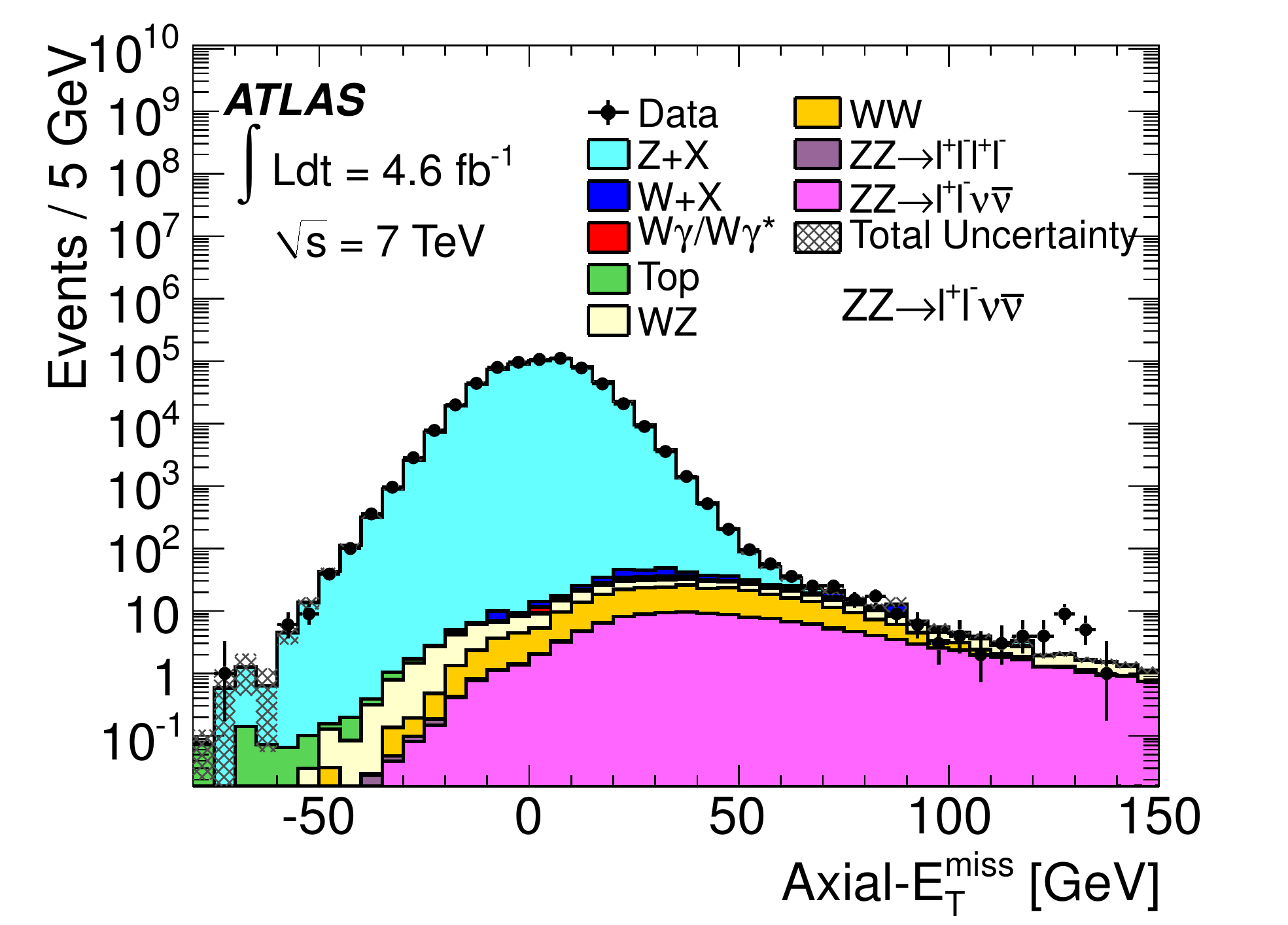}
}
\subfigure[]{
\includegraphics[width=0.47\textwidth]{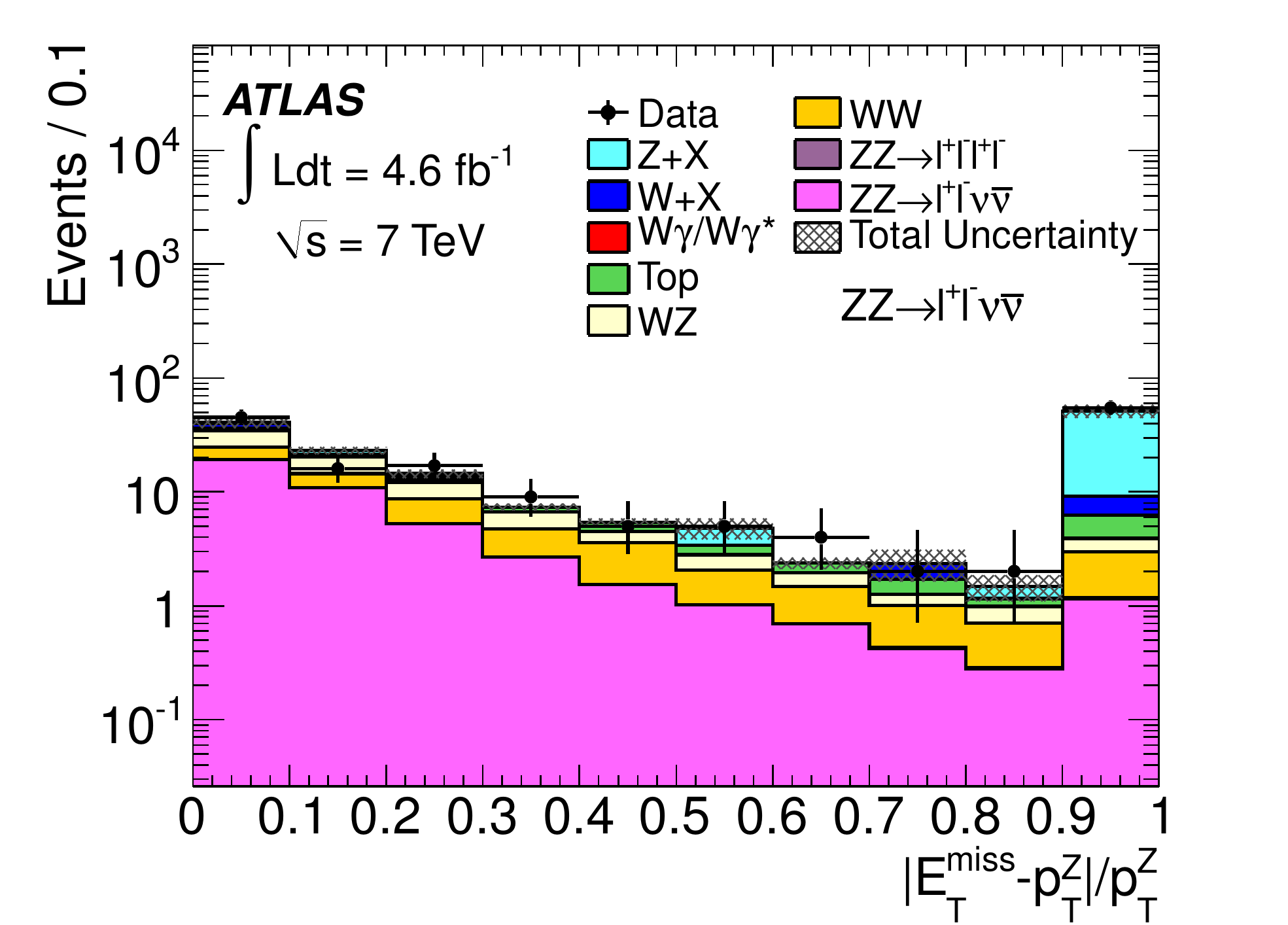}
}
\caption{\label{fig:nminusone_llvv}For \llvv\ candidates in all channels 
          figure (a) shows the axial-$\met$ after all selection requirements, except for the axial-$\met$, and 
          figure (b) shows the fractional \pT\ difference between $\met$ and $p^{Z}_{\rm{T}}$ after all selection requirements, except for the fractional 
          \pT\ difference (the last bin also contains events with fractional \pT\ difference greater than 1). In all plots, the points are data
          and the stacked histograms show the signal prediction from
          simulation. The shaded band shows the combined statistical and systematic uncertainties.
}
\end{figure}

\begin{figure}[htbp] 
\begin{center} 
 \subfigure[]{ 
 \includegraphics[width=0.47\textwidth]{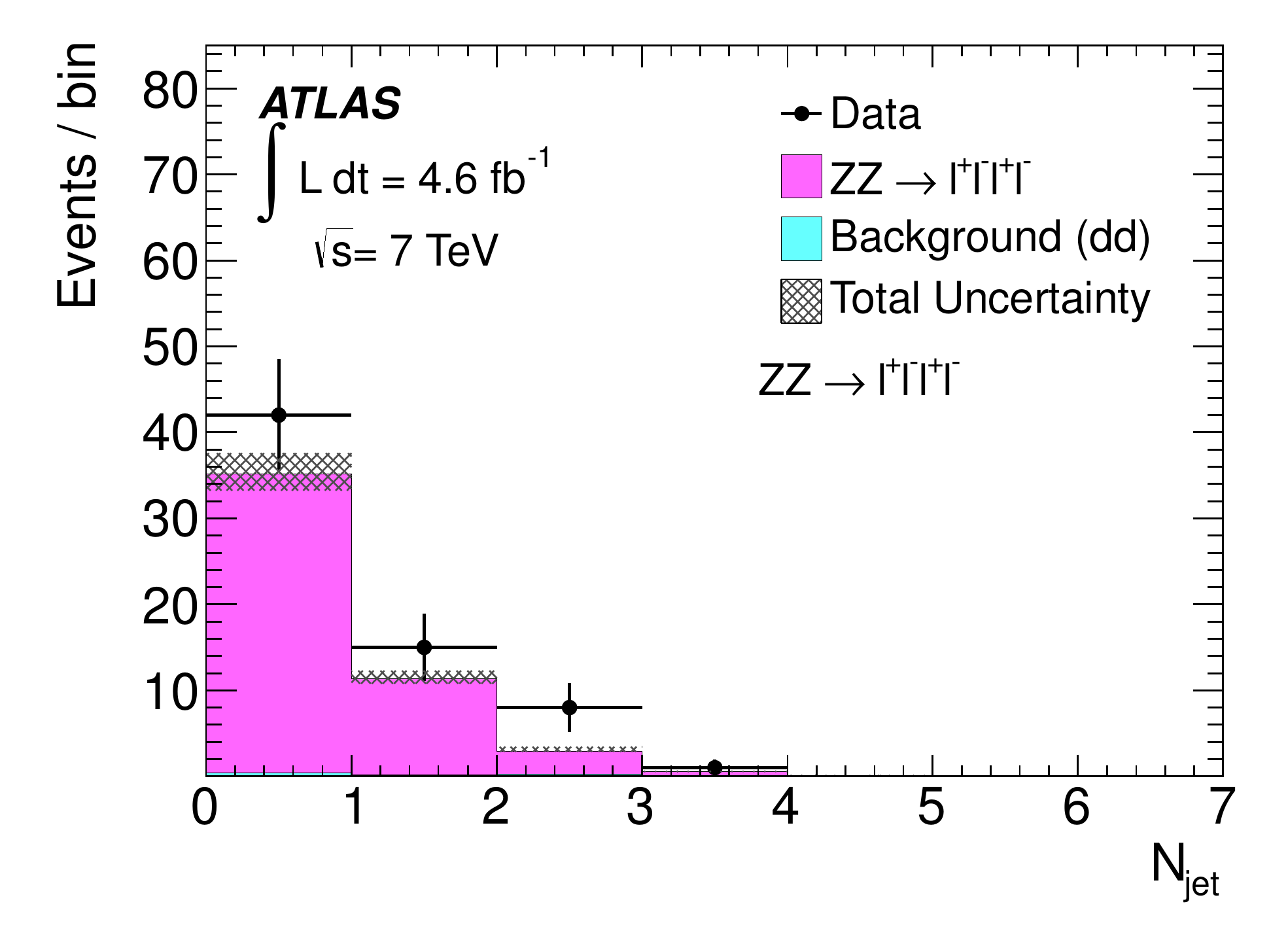} 
 } 
 \subfigure[]{ 
 \includegraphics[width=0.47\textwidth]{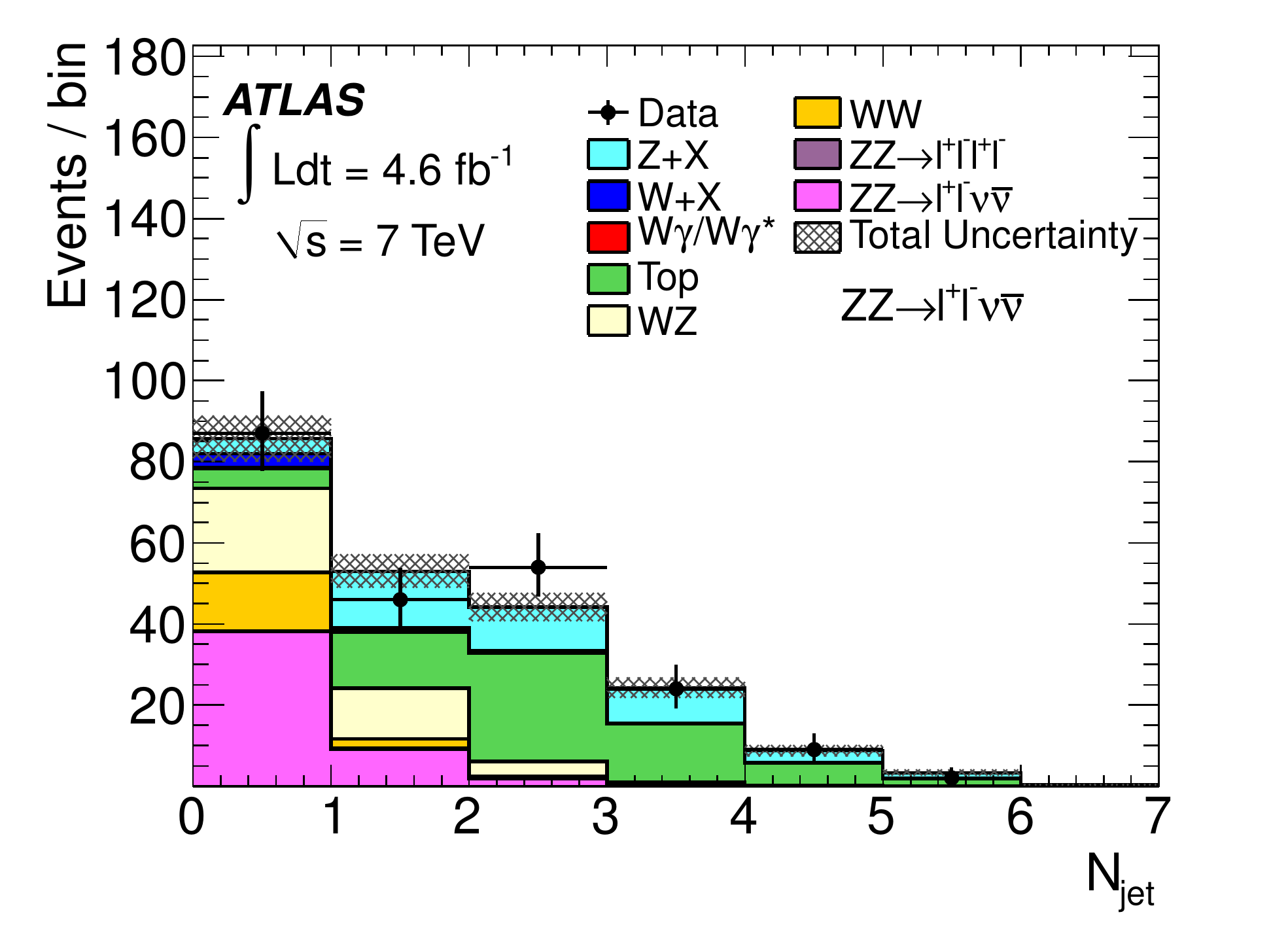} 
 } 
   \caption{\label{fig:kindists_jet}(a) Jet multiplicity for the \zzllll\ selection and (b) jet multiplicity for  
            the \zzllvv\ selection (with all selections applied but the jet veto). 
           The points represent the observed data.
           In (a) the \zzllll\ background is normalized to the data-driven (dd) estimate, 
           while in (b) the histograms show the prediction from simulation.
           The shaded band shows the combined statistical and systematic uncertainty on the prediction. }
 \end{center}
 \end{figure}

\section{Signal acceptance}\label{sec:SignalAcceptance}
        The $Z$ boson decays to hadrons, neutrinos and charged leptons with branching 
fractions of 69.9\%, 20.0\% and 10.1\%, respectively~\cite{PDG}.  
The two $ZZ$ decay channels considered in this paper, \zzllll\  and  \zzllvv, have 
branching fractions of 0.45\% and 2.69\%, respectively\footnote{The quoted branching fraction 
to four charged leptons is for the case where both $Z$ bosons are within the mass window, 
so that the $\gamma^{*}$ contribution can be neglected.}, where 
decays involving $\tau$ leptons are not included in these branching fractions. 
Some of the $ZZ$ decays produce one or more charged leptons which pass through the uninstrumented regions of 
the detector, and as such cannot be reconstructed. In order to measure the total $ZZ$ 
cross section, the measured decays are extrapolated to non-measured parts of the phase-space;
this results in the measurement being more dependent on theory predictions. 
Consequently,  two types of cross sections are measured: fiducial and total. The fiducial cross section 
is the cross section measured within a restricted phase space, and the total cross section is 
the cross section extrapolated to the total phase space.

The total cross section calculation depends on the choice of \Z\ mass range. 
The cross section 
is calculated using the $Z$ boson natural width rather than the zero-width approximation, and includes 
the mass window requirement (66 to 116 GeV) to remove most of the $\gamma^{*}$ contamination. The ratio 
of the total cross section calculated with both \Z\ bosons within the mass window to the total cross section calculated using the 
zero-width approximation is 0.953, as the mass window requirement removes some of the \Z\ bosons 
in the tails of the mass distribution.

\subsection{Fiducial region definitions}

The fiducial cross section is restricted to a region
which is constructed to closely match the instrumented region and
the event selection; for simplicity, only the most inclusive requirements on 
the lepton $\eta$ and \pt\ are used for the definition of the fiducial phase space.
The fiducial cross section $\sigma_{ZZ}^{\rm fid}$ is
calculated as:

\begin{equation}
\sigma_{ZZ}^{\rm fid} = \frac{N_{\rm obs} - N_{\rm
bkg}}{C_{ZZ} \times \mathcal{L}}
\end{equation}

\noindent which depends on a correction factor given by the number of
simulated \ZZS\ events 
which satisfy the full event selection divided by the number of \ZZS\ events
generated in the fiducial region, $C_{ZZ}$; the integrated luminosity, $\mathcal{L}$; the number of selected events, $N_{\rm obs}$; and the amount of estimated background, $N_{\rm bkg}$. 
For the calculation of $C_{ZZ}$, final states including pairs of oppositely-charged leptons produced from 
decays of $Z\to\tau^{+}\tau^{-}\to\ll\nu\bar{\nu}\nu\bar{\nu}$ are included in the number of 
selected events (numerator) since those decays 
have an identical final state to the signal and are not subtracted as
background but are excluded from the fiducial region (denominator)
because the fiducial regions are defined only with \ZZS\ decays directly to
electrons, muons or neutrinos, depending on the channel. 
The contribution from such $\tau$ decays is estimated from Monte Carlo simulation to be
$<0.1$~\% for the \zzllvv\ selection, 0.24$\pm$0.01\% 
for the \zzllll\ selection and 1.73$\pm$0.04\% for the \zzsllll\
selection.
Fiducial requirements are applied at generator level. To reduce the dependence on QED radiation, {\emph {dressed}} 
leptons are always used, for which the lepton four-momentum is summed with the four-momentum of 
all photons within $\Delta{R}=0.1$. 

The \zzllll\ fiducial region is defined using the following requirements: 
(i) two pairs of same-flavour opposite-sign electrons or muons, with each lepton satisfying $p_{\rm T}^{\ell} > 7$ GeV, 
    $|\eta^{\ell}|<3.16$ and at least a distance $\Delta R = 0.2$ from any other
selected lepton, i.e., $\Delta R(\ell_1,\ell_2)>0.2$, and (ii) both dilepton invariant masses within the \Z\ mass window.
 A \zzsllll\ fiducial region is defined with the same criteria as in the \zzllll\ case, except that one dilepton invariant mass requirement is relaxed to be greater than 20 GeV.

The  $ZZ\to\ll \nu\bar{\nu}$ fiducial region is defined by requiring: 
(i)  two same-flavour opposite-sign electrons or muons, each with $p_{\rm
    T}^{\ell} > 20$ GeV, $|\eta^{\ell}|<2.5$, with $\Delta R(\ell^+,\ell^-)>0.3$,
(ii) dilepton invariant mass close to the $Z$ boson mass:
  $76 < m_{\ll} < 106$ GeV,
(iii) dineutrino invariant mass close to the $Z$ boson mass:
  $66 < m_{\nu\bar{\nu}} < 116$ GeV,
(iv) no jet with $p^j_{\rm T} >$ 25 GeV and
  $|\eta^j|<4.5$,
and (v) $(|p_{\rm T}^{\nu\bar{\nu}} - p_{\rm T}^Z|)/ p_{\rm T}^Z < 0.4$
  and $-\vec{p}_{\mathrm{T}}^{\nu\bar{\nu}} \cdot \vec{p}^Z / p_{\mathrm T}^Z>75$ GeV.
  Jets are defined at generator level using the same jet algorithm as
  used in reconstructed events and including all final state particles after parton 
  showering and hadronization.

Fiducial cross sections are calculated using the \zzllll, \zzsllll\ and \zzllvv\ 
selections, integrated over the corresponding full fiducial phase space volumes.
For the \zzllll\ and  \zzllvv\ selections the differential fiducial cross sections are derived in bins of the leading 
$p_{\rm T}^{Z}$, $\Delta\phi(\ell^+,\ell^-)$ and the mass of the \zzllll\ system or the transverse mass 
of the \zzllvv\ system.

The correction factor, $C_{ZZ}$, is determined from 
Monte Carlo simulations 
(\powhegbox\ for the \zzllvv\ channel and \powhegbox\ and \ggtwozz\ for the
\zzSllll\ channel), after applying data-driven corrections as
described in section~\ref{sec:MCsample}.
For the \zzllll\ (\zzsllll) selection it is 0.43 (0.41) for $e^{+}e^{-}e^{+}e^{-}$, 0.68 (0.69) for $\mu^{+}\mu^{-}\mu^{+}\mu^{-}$ and 0.55 (0.53) for $e^{+}e^{-}\mu^{+}\mu^{-}$ events.
For the \zzllvv\ selection the correction factor is 0.63 for $e^{+}e^{-}\nu\bar{\nu}$ and 0.76 for $\mu^{+}\mu^{-}\nu\bar{\nu}$ events.
The correction factors combining all lepton categories within the fiducial region are given in table~\ref{tab:czztable} for the three event selections in both decay channels.

\begin{table}[htbp]
     \centering
     \begin{tabular}{lc}
     \hline
     Selection & $C_{ZZ}$ \\
     \hline
     \zzllll\ & 0.552 $\pm$ 0.002 $\pm$ 0.021 \\
     \hline
     \zzsllll\ & 0.542 $\pm$ 0.002 $\pm$ 0.022 \\
     \hline
     \zzllvv\ & 0.679 $\pm$ 0.004 $\pm$ 0.014 \\
     \hline
     \end{tabular}
      \caption{\label{tab:czztable} Correction factors $C_{ZZ}$ for each production and decay channel. 
          The first uncertainty is statistical while the second is systematic. 
      }
\end{table}

\subsection{Extrapolation to the total phase space}

The total \ZZ\ cross section is measured using the \zzllll\ and \zzllvv\ selections. 
The total cross section is calculated using the fiducial acceptance, $A_{ZZ}$ 
(the fraction of $ZZ$ events with $Z$ bosons in the \Z\ mass window that fall into the 
fiducial region) and the branching fraction, BF: 

\begin{equation}
\sigma_{ZZ}^{\rm total} = \frac{N_{\rm obs} - N_{\rm
bkg}}{A_{ZZ} \times C_{ZZ} \times \mathcal{L} \times {\rm BF}}
\end{equation}

The fiducial acceptances $A_{ZZ}$ are estimated from Monte Carlo simulation, using
\powhegbox\ for the \zzllvv\ channel and \powhegbox\ and \ggtwozz\ for the
\zzllll\ channel. The fiducial acceptance of the \zzllvv\ channel is much more 
constrained than the \zzllll\ channel in order to reduce background. Values are given in table~\ref{tab:azztable}. 

\begin{table}[htbp]
     \centering
     \begin{tabular}{lc}
     \hline
     Selection & $A_{ZZ}$ \\
     \hline
     \zzllll\ & 0.804 $\pm$ 0.001 $\pm$ 0.010 \\
     \hline
     \zzllvv\ & 0.081 $\pm$ 0.001 $\pm$ 0.004 \\
     \hline
     \end{tabular}
      \caption{\label{tab:azztable}Acceptance $A_{ZZ}$ for the two decay channels used for the measurement of the total \ZZ\ production cross section. 
          The first uncertainty is statistical while the second is systematic. 
      }
\end{table}

\subsection{Systematic uncertainties}

Table~\ref{tab:sys} summarizes the systematic uncertainties on $C_{ZZ}$ and $A_{ZZ}$.  
For $C_{ZZ}$ in the \zzSllll\ selections, 
the dominant systematic uncertainties arise from the lepton reconstruction efficiency, 
the efficiency of the isolation and impact parameter requirements, and the
differences in $C_{ZZ}$ estimated by \sherpa\ and \powhegbox; 
uncertainties on the trigger efficiency and the lepton energy scale and resolution are small.
In the \zzllvv\ channel the dominant $C_{ZZ}$ uncertainties are from uncertainties on the
lepton reconstruction efficiency, the lepton energy scale and resolution, and
the missing transverse momentum modelling and jet veto uncertainty; uncertainties
on the trigger efficiency and due to differences in $C_{ZZ}$ estimated by
\sherpa\ and \powhegbox\ also contribute. 

The uncertainties on $C_{ZZ}$ from the reconstruction efficiency, energy scale and
resolution, isolation and impact parameter requirements and trigger efficiency 
are estimated by varying the data-driven correction factors applied to simulation 
by their systematic and statistical uncertainties. 
The systematic uncertainties on events with extended leptons used in the \zzSllll\ channel are slightly higher than in events without them; nevertheless, since
their relative contribution is small, the effect on the uncertainty of the combined channels is negligible. 
The generator systematic uncertainty for $C_{ZZ}$ accounts for the effect of choosing a different renormalization and factorization scale and PDF set.

For $A_{ZZ}$, the systematic uncertainties 
are due to theoretical uncertainties which come from the PDFs, 
the choice of the renormalization and factorization scales, 
the modelling of the contribution from $gg$ initial states and
the parton shower model, as given in table~\ref{tab:sys}. 
For the \zzllvv\ channel, uncertainties in the efficiency of the jet veto are also taken into account through the calculation of a scale factor; 
the ratio of the jet veto efficiency in data to that in MC simulation is taken from a sample of single $Z$ events and then applied to $ZZ$ 
events~\cite{WWPaper}. 
The systematic uncertainties due to the PDFs and scales are evaluated 
with MCFM by taking the difference between the $A_{ZZ}$ obtained using the CT10
and MSTW2008 PDF sets, as well as using the 44 CT10 error sets, and by shifting the
factorization and renormalization scales up and down by a factor of two.
An additional uncertainty is assigned to account for the effect of different modelling at the generator level. 
Since the \zzsllll\ measurement is not used for the total cross section, 
its $A_{ZZ}$ acceptance is irrelevant and only uncertainty values related to $C_{ZZ}$ are given.

The uncertainty on the integrated luminosity is 3.9\%~\cite{bib:ATLASLumi2010}. The uncertainty on the background estimates is discussed in the following sections.

\begin{table}[htbp]
\centering
\begin{tabular}{lccc}
\hline
Source & \zzllll\ & \zzsllll\ & \zzllvv\   \\\hline
\multicolumn{4}{c}{} \\
\multicolumn{4}{c}{$C_{ZZ}$} \\\hline
Lepton efficiency        & 3.0\% & 3.1\% & 1.3\% \\
Lepton energy/momentum   & 0.2\% & 0.3\% & 1.1\% \\
Lepton isolation and impact parameter  & 1.9\% & 2.0\% & 0.6\% \\
Jet+\missET\ modelling   & -- & -- & 0.8\% \\
Jet veto                 & -- & -- & 0.9\% \\
Trigger efficiency       & 0.2\% & 0.2\% & 0.4\% \\
PDF and scale            & 1.6\% & 1.5\% & 0.4\% \\ 
\hline
\multicolumn{4}{c}{} \\
\multicolumn{4}{c}{$A_{ZZ}$} \\
\hline
Jet veto                 & -- & -- & 2.3\% \\
PDF and scale            & 0.6\% & --    & 1.9\% \\
Generator modelling and parton shower & 1.1\% & --    & 4.6\% \\ 
\hline
\end{tabular}
\caption{\label{tab:sys}Summary of systematic uncertainties, as relative percentages 
  of the correction factor $C_{ZZ}$ or the acceptance of the fiducial region
  $A_{ZZ}$. Dashes indicate uncertainties which are not relevant. }
\end{table}

\section{Background estimation}\label{sec:Background}
        \subsection{\zzSllll\ background}\label{sec:Background4l}
        Background to the \zzSllll\ 
signal originates from events with a \Z\ (or $W$) boson decaying to leptons
accompanied by additional jets or photons ($W/Z+X$), from top-quark production and from other diboson
final states.
Such events may contain electrons or muons from the decay of heavy-flavoured hadrons, muons from in-flight 
decay of pions and kaons, or jets and photons misidentified as electrons. 
The majority of these background leptons are rejected by the isolation requirements. 

The background estimate follows a data-driven method 
in which a sample of events containing 
three leptons satisfying all selection criteria plus one `lepton-like jet' is identified;
such events are denoted as $\ell \ell \ell j$.
For muons, the lepton-like
jets are muon candidates that fail the isolation requirement or fail
the impact parameter requirement but not both. For electrons with $|\eta| <
2.47$, the lepton-like
jets are clusters in the electromagnetic calorimeter matched to inner
detector tracks that fail either the full electron selection or the
isolation requirement but not both. 
For electrons with  $|\eta| > 2.5$, the lepton-like jets are electromagnetic
clusters that are reconstructed as electrons but fail the tight identification
requirements. The events
are otherwise required to satisfy the full event selection, treating the lepton-like jet as if it
were a fully identified lepton. 
The background is then estimated by weighting the $\ell \ell \ell j$ events 
by a measured factor $f$, 
which is the ratio of the probability for a non-lepton to 
satisfy the full lepton selection criteria to the probability of satisfying the lepton-like jet criteria.
The background in which two selected leptons originate from jets is treated similarly, by 
identifying a data sample with two leptons and two lepton-like jets;
such events are denoted as $\ell \ell j j$.
The total number of expected background \llll\ events, $N(\rm{BG})$, is calculated as:
\begin{linenomath}
\begin{equation}
N({\rm BG}) = [ N(\ell\ell\ell j) - N(\ZZ) ] \times f - N(\ell\ell jj)\times f^2
\label{eq:zz4l_bkg_formula}
\end{equation}
\end{linenomath}
where double counting from $\ell \ell \ell j$ and $\ell \ell j j$ events is accounted for, and the term $N(\ZZ)$ is a Monte Carlo estimate correcting for contributions from signal \zzSllll\ events having a real lepton that is classified as a lepton-like jet (the equivalent correction to the term $N(\ell\ell jj)$ is negligible).

The factor $f$ is measured in a sample of data selected with
 single-lepton triggers which contain a $Z$ boson candidate: 
a pair of isolated same-flavour opposite-sign electrons or muons. 
In these selected events, $f$ is measured,
using the lepton and lepton-like jet candidates not assigned to the \Z\ boson,
as the ratio of the number of selected leptons to the number of lepton-like jets,
after correcting for expected true lepton contributions from $WZ$ and $ZZ$ events using simulation.
Independent values as a function of the \eta\ and \pT\  of the lepton-like jet are measured, which are then combined assuming they are uncorrelated.
The factor $f$ is found to vary 
from $0.33\pm0.01$ ($0.26\pm0.02$) below $\pT=10$ \GeV\ to  $0.09\pm0.02$ ($0.46\pm0.20$) above $\pT=50$ \GeV\ for electrons (muons).  
The quoted uncertainties are statistical.
Then, with the same procedure, a value for $f$ is also derived using the
simulated samples of background processes. 
The difference between the value of $f$ derived in data and in simulation is assigned as a systematic uncertainty on $f$. 
The statistical and systematic uncertainties are then added in quadrature to derive a combined uncertainty on $f$, 
which varies as a function of \pT\
from 14\% (19\%) below 10 \GeV\ to 22\% (51\%) above 50 \GeV\ for electrons (muons).
For the muons, the total uncertainty on $f$ is dominated by its statistical uncertainty. 
The background estimates for the \zzllll\ and \zzsllll\ selections are
$0.9^{+1.1 }_{-0.9 }\rm{(stat.)} \pm 0.7 \rm(syst.)$ and 
$9.1  \pm 2.3  \rm{(stat.)} \pm 1.3 \rm{(syst.)}$ events, respectively, as  shown in tables~\ref{tab:zzfakes} and~\ref{tab:zzsfakes}.
The statistical uncertainty on the background estimate comes from the statistical uncertainty on the numbers of $\ell \ell \ell j$, $\ell \ell j j$ and  \zzSllll\ events used in eq.~\ref{eq:zz4l_bkg_formula}.
The systematic uncertainty results from the combined uncertainty on $f$.
In cases where the overall estimate is negative, the background
estimate is described using a truncated Gaussian with mean at zero and 
standard deviation equal to the estimated statistical and systematic uncertainties added in quadrature. 

The extra background induced by the use of the extended leptons in the \zzSllll\ channel is estimated to be negligible in the \zzllll\ selection, and about 20\% 
(2 events out of the 9.1 estimated, compared to a signal gain of about 10.6 events out of the 64.4 expected) in the \zzsllll\ selection.

The background is also estimated purely from the simulated samples of background processes, and  is predicted to be 
1.5 $\pm$ 0.4  events for the \zzllll\ selection and 8.3 $\pm$ 1.3 events for the \zzsllll\ selection, with uncertainties being statistical only. 
These estimates compare well with the data-driven results given in tables~\ref{tab:zzfakes} and~\ref{tab:zzsfakes}.
According to the estimate from simulation, the dominant source of background is $Z$+jets events, with only about a 10\% to 20\% contribution from other diboson channels ($WZ$ and $WW$), and a negligible contribution from events with top quarks.

\begin{table}
\begin{center}
\begin{tabular}{lrrrr}
\hline 
& $e^{+}e^{-}e^{+}e^{-}$ & $\mu^{+}\mu^{-}\mu^{+}\mu^{-}$ & $e^{+}e^{-}\mu^{+}\mu^{-}$ & \llll \\ \hline
$(+)\ N(\ell\ell\ell j) \times f$     & $1.63 \pm 0.34$ & $0.21 \pm 0.21 $         & $1.84 \pm 0.40 $ & $3.67 \pm 0.57 $ \\
$(-) N(ZZ) \times f$ & $0.17 \pm 0.13$ & $0.12 ^{+0.20} _{-0.12}$ & $0.34 \pm 0.21 $ & $0.63 \pm 0.32 $ \\
$(-) N(\ell\ell jj) \times f^2 $   & $0.96 \pm 0.10$ & $0.33 \pm 0.16 $         & $0.83 \pm 0.09 $ & $2.12 \pm 0.21 $ \\
Background estimate, $N({\rm BG})$  & $0.5^{+0.6} _{-0.5} \rm{(stat.)}$ & $ <0.64$ & $0.7 \pm 0.7 \rm{(stat.)}$ & $0.9  ^{+1.1 } _{-0.9 }\rm{(stat.)}$  \\
                                            & $     \pm 0.3 \rm{(syst.)}$ &                   &   $\pm0.6$ (syst.) & $    \pm 0.7 \rm{(syst.)}$ \\
\hline
\end{tabular}
\end{center}
\caption{Expected number of background events for the \zzllll\ selection in \lumi\ of data, for the individual decay modes (columns 2, 3 and 4) and for their combination (last column). If the central value of the estimate is negative, the upper bound on the number of events in that channel is derived as detailed in section 5.1.}
\label{tab:zzfakes}
\end{table}

\begin{table}
\begin{center}
\begin{tabular}{lrrrr}
\hline
& $e^{+}e^{-}e^{+}e^{-}$ & $\mu^{+}\mu^{-}\mu^{+}\mu^{-}$ & $e^{+}e^{-}\mu^{+}\mu^{-}$ & \llll \\ \hline
$(+)\ N(\ell\ell\ell j) \times f$     & $8.85 \pm 0.98$   & $0.21 \pm 0.21 $           & $10.63 \pm 1.06 $ & $19.70 \pm 1.46 $ \\
$(-) N(ZZ) \times f$ & $0.29 \pm 0.18$   & $0.20  ^{+0.25} _{-0.20}$   & $0.56 \pm 0.28 $ & $1.05 \pm 0.42 $ \\
$(-) N(\ell\ell jj) \times f^2 $   & $4.24 \pm 0.23$   & $1.10 \pm 0.31 $            & $4.24 \pm 0.23 $ & $9.58 \pm 0.45 $ \\
Background estimate, $N({\rm BG})$  & $4.3  \pm 1.4 \rm{(stat.)}$ & $ <0.91$ & $5.8  \pm 1.6  \rm{(stat.)}$ & $9.1  \pm 2.3  \rm{(stat.)}$  \\
                                            & $     \pm 0.6  \rm{(syst.)}$ &          &        $\pm0.9 $ (syst.) & $    \pm 1.3  \rm{(syst.)}$ \\
\hline
\end{tabular}
\end{center}
\caption{Expected number of background events for the \zzsllll\ selection in \lumi\ of data, for the individual decay modes (columns 2, 3 and 4) and for their combination (last column). If the central value of the estimate is negative, the upper bound on the number of events in that channel is derived as detailed in section 5.1.}
\label{tab:zzsfakes}
\end{table}

Differential background distributions are determined by first
deriving the shape of the distributions from the background MC samples.
This is achieved by selecting events where one \Z\ candidate is required to satisfy the nominal lepton selection, while the other \Z\ candidate is formed by leptons satisfying relaxed criteria for the
isolation requirements and transverse impact parameter significance.
The shape determined in this way  is then scaled such that the total number of events in the distribution is equal to the data-driven background estimate shown in tables~\ref{tab:zzfakes} and~\ref{tab:zzsfakes}.

        \subsection{\zzllvv\ background}
        There are several sources of background to the \zzllvv\ 
channel.  Processes such as $t\bar{t}$, $WW$, $Wt$ or
\ztt\ production give two true isolated leptons with
missing transverse momentum.  Diboson $WZ$ events in which both bosons
decay leptonically have three charged leptons, but if one lepton from a $W$ or $Z$ boson decay is not
identified, the event has the same signature as the signal. Production of a
$Z$ boson in association with jets gives two isolated leptons from the
$Z$ boson decay and may have missing transverse momentum if the jet momenta
are mismeasured.  Finally, production of a $W$ boson in association
with jets or photons may satisfy the selection requirements when one of the 
jets or photons 
is misidentified as an isolated lepton. All of the backgrounds are 
measured with data-driven techniques except for $WZ$ and $W\gamma$. 
The total background is estimated to be $46.9\pm4.8\pm1.9$ events as summarized in table~\ref{tab:2l2nu}.

\begin{table}
\centering
\begin{tabular}{lrrr}
\hline
Process & $e^+e^- $\missET & $\mu^+\mu^- $\missET & $\ell^+\ell^- $\missET \\ 
\hline
$t\bar{t}$, $Wt$, $WW$, \ztt  & $8.5 \pm 2.1 \pm 0.5$ & $10.6\pm 2.6 \pm 0.6$ & $19.1\pm 2.3 \pm 1.0$ \\
$WZ$                          & $8.9 \pm 0.5 \pm 0.4$ & $11.9\pm 0.5 \pm 0.3$ & $20.8\pm 0.7 \pm 0.5$ \\
$Z\ra\mu^+\mu^-, e^+e^-$+jets & $2.6 \pm 0.7 \pm 1.0$ & $2.7 \pm 0.8 \pm 1.2$ & $5.3 \pm 1.1 \pm 1.6$ \\
$W+$ jets                     & $0.7 \pm 0.3 \pm 0.3$ & $0.7 \pm 0.2 \pm 0.2$ & $1.5 \pm 0.4 \pm 0.4$ \\
$W\gamma$                     & $0.1 \pm 0.1 \pm 0.0$ & $0.2 \pm 0.1 \pm 0.0$ & $0.3 \pm 0.1 \pm 0.0$ \\
\hline
Total                         & $20.8\pm 2.3 \pm 1.2$ & $26.1\pm 2.8 \pm 1.4$ & $46.9\pm 4.8 \pm 1.9$ \\ 
\hline
\end{tabular}

\caption{\label{tab:2l2nu}Expected number of background events to the $ZZ\ra
         \ell^+\ell^-\nu\bar{\nu}$ channel in \lumi\ of data, for the individual decay modes 
         (columns 2 and 3) and for their combination (last column). The first uncertainty is 
         statistical while the second is systematic.}
\end{table}

\subsubsection{Backgrounds from $t\bar{t}$, $Wt$, $WW$ and
\ztt }

The contributions from $t\bar{t}$, $Wt$, $WW$ and \ztt\ processes are 
measured  by extrapolating from a control sample of events with one electron 
and one muon (instead of two electrons or two muons), which otherwise satisfy the full \zzllvv\ selection. This 
sample is free from signal events. The extrapolation from the $e\mu$ channel to the 
$ee$ or $\mu\mu$ channel uses the relative branching fractions ($2:1:1$ for $e\mu:ee:\mu\mu$) as well as the ratio of 
the efficiencies $\epsilon_{ee}$ or $\epsilon_{\mu\mu}$ of the $ee$ or $\mu\mu$ selections to the efficiency 
$\epsilon_{e\mu}$ of the $e\mu$
selection, which differs from unity due to differences in the electron
and muon efficiencies. 

For the electron channel, this is represented by 
the equation: 
\begin{equation}
N_{ee}^{\rm bkg} =  ( N^{\rm data}_{e\mu} - N^{\rm sim}_{e\mu} )
\times \frac{1}{2} \times \frac{\epsilon_{ee}}{\epsilon_{e\mu}}
\end{equation}
\noindent where  $N^{\rm data}_{e\mu}$ is the number of observed $e\mu$ events 
and $N^{\rm sim}_{e\mu}$ is the number
expected events from processes other than $t\bar{t}$, $Wt$, $WW$ and
\ztt\ ($WZ$, $ZZ$, $W+$jet, $Z+$jet and $W/\gamma$).  
Therefore, $( N^{\rm data}_{e\mu} - N^{\rm sim}_{e\mu} )$ is the estimate of $t\bar{t}$, $Wt$, $WW$ and \ztt\ production in the 
control sample. The efficiency correction factor, $\epsilon_{ee}/\epsilon_{e\mu}$,
corrects for the difference between electron and muon efficiency.
The efficiency correction factor is measured in data using reconstructed $Z\rightarrow\ll$ events, as 

\begin{equation}
\frac{\epsilon_{ee}}{\epsilon_{e\mu}} =
\frac{\epsilon_e^2}{\epsilon_e\epsilon_\mu} = \frac{\epsilon_{e}}{\epsilon_{\mu}} =\sqrt{\frac{ N^{\rm
data}_{ee}}{N^{\rm data}_{\mu\mu}}}
\end{equation}

\noindent
where $N^{\rm data}_{ee}$ and $N^{\rm data}_{\mu\mu}$ are the number
of observed $ee$ or $\mu\mu$ events in the $Z$ boson mass window, respectively, after all lepton selection requirements 
and the $Z$ boson mass window requirement are applied. 
A parallel argument gives $N_{\mu\mu}^{\rm bkg}$.
This procedure is repeated in bins of $p_{\rm{T}}^{Z}$ in order to obtain the \pT\ distribution of the 
$t\bar{t}$, $Wt$, $WW$ and \ztt\ background.

The dominant uncertainty is statistical (25\%), due to the limited number of
events in the control samples. Additional uncertainties are due to
systematic uncertainties in the normalization of the simulated samples used to correct the
$e\mu$ contribution (5.5\%) and the systematic uncertainty in the efficiency correction factor (4.5\%).

\subsubsection{Background from $WZ$ production with leptonic decays}

Events from leptonic $WZ$ decays may result in an $\ll$\missET\ signature when one lepton from the 
$W$ or $Z$ boson is not reconstructed.
The contribution from this process is estimated using the simulated
samples described in section 2.1. The estimate is checked using a control region with three high-\pT\ isolated
leptons. The two dominant processes that contribute to this control region are $WZ$ and $Z+$jets production, 
where the $WZ$ boson pair 
decays to three leptons and a neutrino and the $Z+$jets contribution has two real leptons from the $Z$ decay 
and a misidentified lepton from the jet.
The technique used to estimate the background in the \zzSllll\ channel is also used to normalize 
the contribution from $Z+$jets in the three-lepton control region. The $WZ$ Monte Carlo expectation 
is consistent with the data. The systematic uncertainties are estimated in the same way as for 
signal Monte Carlo events.

\subsubsection{Background from $Z$ bosons with associated jets}

Occasionally events with a $Z$ boson produced in association with jets 
may have large amounts of missing transverse momentum due to 
mismeasurement of the momenta of the jets. This background 
is estimated using events with a high-\pT\ photon and 
jets as a template, since the mechanism for large missing transverse momentum is the
same as in $Z$+jets events.  The events are reweighted such that the photon
$E_{\rm T}$ matches the observed $Z$ boson \pT\ and are normalized to the
observed $Z$ + jets yield.
The procedure is repeated in bins of $p_{\rm{T}}^{Z}$ in order to obtain the \pT\
distribution of  the $Z+$jets backgrounds.
The largest systematic uncertainty is due to the subtraction of
$W\gamma$, $Z\gamma$, $t\bar{t}$ and $W\rightarrow e\nu$ contributions to
the $\gamma$+jets sample, which is 33\% in the $ee$ channel and 37\% in the 
$\mu\mu$ channel.

\subsubsection{Background from events with a misidentified lepton}

A small contribution to the selected sample  is due to events in which one of the two leptons comes
from the decay of a $W$ or $Z$ boson (called `real' below) and the second is a `fake',
corresponding both non-prompt leptons and misidentified $\pi^0$ mesons or
conversions. 

The dominant fake-muon mechanism is the decay of heavy-flavoured hadrons, 
in which a muon survives the isolation
requirements. In the case of electrons, the three mechanisms are heavy-flavour hadron 
decay, light-flavour jets with a leading $\pi^0$ overlapping
with a charged particle, and conversion of photons.  Processes that contribute 
are top-quark pair production, production of $W$ bosons in association with
jets and multi-jet production.

The `matrix method'~\cite{bib:2012xh} is applied to estimate the fraction
of events in the signal regions that contain at least one fake
lepton.  The method measures the number of fake leptons in
background-dominated control regions and extrapolates to the $ZZ$
selection region using factors measured in data. 
The shape of the background is provided by taking the background as uniformly 
distributed among the bins and treating each bin as statistically uncorrelated. 
The dominant systematic uncertainty is due to the uncertainty on the
extrapolation factors and the limited numbers of events in the control samples, giving a 
total uncertainty of 63\% and 44\% in the $ee$ and $\mu\mu$ channels, respectively.

\section{Results}\label{sec:Results}
        
Three types of measurements are presented: 
\begin{itemize}
\item integrated fiducial and total \ZZ\ cross sections; 
\item differential cross sections normalized to the overall measured
cross sections for the $p_{\rm T}^{Z}$ and $\Delta\phi(\ell^+,\ell^-)$ of the leading $Z$ boson, 
and the mass (transverse mass\footnote{$m_{\rm T}^2=\left( \sqrt{ (m^{Z})^2 + (p_{\rm T}^{Z})^2 } 
+ \sqrt{ (m^{Z})^2 + (E_{\mathrm{T}}^{\mathrm{miss}})^2 } \right)^{2}
- \left( \vec{p}_{\rm T}^{Z} + \vec{E}_{\mathrm{T}}^{\mathrm{miss}} \right)^2$}) 
of the \ZZ\ system for the \zzllll\ (\zzllvv) selection; and
\item limits on the anomalous nTGCs.  
\end{itemize}

        \subsection{Cross section measurements}\label{sec:xSec}
        The expected and observed event yields after applying all selection
criteria are shown in table~\ref{tab:selected_data_MC} for both channels. 
Figure~\ref{fig:kindists_jet} shows the jet multiplicity in selected
\zzllll\ and \zzllvv\ events before the jet veto is applied.
Figures~\ref{fig:kindists_zz} and~\ref{fig:kindists_zzs} show the transverse
momentum and mass of the \ZZ\ system in selected \zzllll\ and \zzsllll\ events
respectively.
Figure~\ref{fig:kindists_zz2l2n} shows the transverse momentum and mass of the
two-charged-lepton system in selected \zzllvv\ events.
The shapes of the distributions are consistent with the predictions from the simulation. 

\begin{table}
\centering
  \begin{tabular}{lcccc}
    \hline
     \zzSllll             & $e^{+}e^{-}e^{+}e^{-}$ & $\mu^{+}\mu^{-}\mu^{+}\mu^{-}$ & $e^{+}e^{-}\mu^{+}\mu^{-}$ & \llll \\
     \hline
Observed \ZZ\ & 16 & 23 & 27 & 66 \\
Observed \ZZs\ & 21 & 30 & 33 & 84 \\
     \hline
Expected \ZZ\ signal &   10.3 $\pm$ 0.1 $\pm$ 1.0 &  16.5 $\pm$ 0.2 $\pm$ 0.9 &  26.7 $\pm$ 0.2 $\pm$ 1.7 &  53.4 $\pm$ 0.3 $\pm$ 3.2 \\
Expected \ZZs\ signal &  12.3 $\pm$ 0.2 $\pm$ 1.2 &  20.5 $\pm$ 0.2 $\pm$ 1.1 &  31.6 $\pm$ 0.3 $\pm$ 2.0 &  64.4 $\pm$ 0.4 $\pm$ 4.0 \\
\hline
Expected \ZZ\ background  & 0.5 $\pm$ 0.6 $\pm$ 0.3 & $<0.6$ & 0.7 $\pm$ 0.7 $\pm$ 0.6 & 0.9 $\pm$ 1.1 $\pm$ 0.7 \\
Expected \ZZs\ background & 4.3 $\pm$ 1.4 $\pm$ 0.6 & $<0.9$ & 5.8 $\pm$ 1.6 $\pm$ 0.9 & 9.1 $\pm$ 2.3 $\pm$ 1.3 \\
    \hline  \\ \hline
 \zzllvv             & $e^{+}e^{-} \MET$ & $\mu^{+}\mu^{-} \MET$ &  & $\ll \MET$ \\
    \hline 
    Observed \ZZ\        & $35$ & $52$ &  & $87$ \\
    \midrule
    Expected \ZZ\  signal      & $17.8\pm0.3\pm1.7$ & $21.6\pm0.3\pm2.0$ &  & $39.3\pm0.4\pm3.7$ \\
    \hline
    Expected \ZZ\  background  & $20.8\pm2.3\pm1.2$ & $26.1\pm2.8\pm1.4$ & & $46.9\pm4.8\pm1.9$ \\
    \hline
  \end{tabular}

  \caption{\label{tab:selected_data_MC}
           Summary of observed \zzllll, \zzsllll\ and \zzllvv\ candidates in the data, total background estimates and expected signal
       for the individual decay modes (columns 2 to 4) and for their combination (last column).
       The quoted uncertainties and limits represent 68\% confidence intervals; the first uncertainty is statistical
           while the second is systematic. The uncertainty on the
       integrated luminosity (3.9\%) 
       is not included. 
          }

\end{table}

\begin{figure}[htbp]
\begin{center}
 \subfigure[]{
 \includegraphics[width=0.47\textwidth]{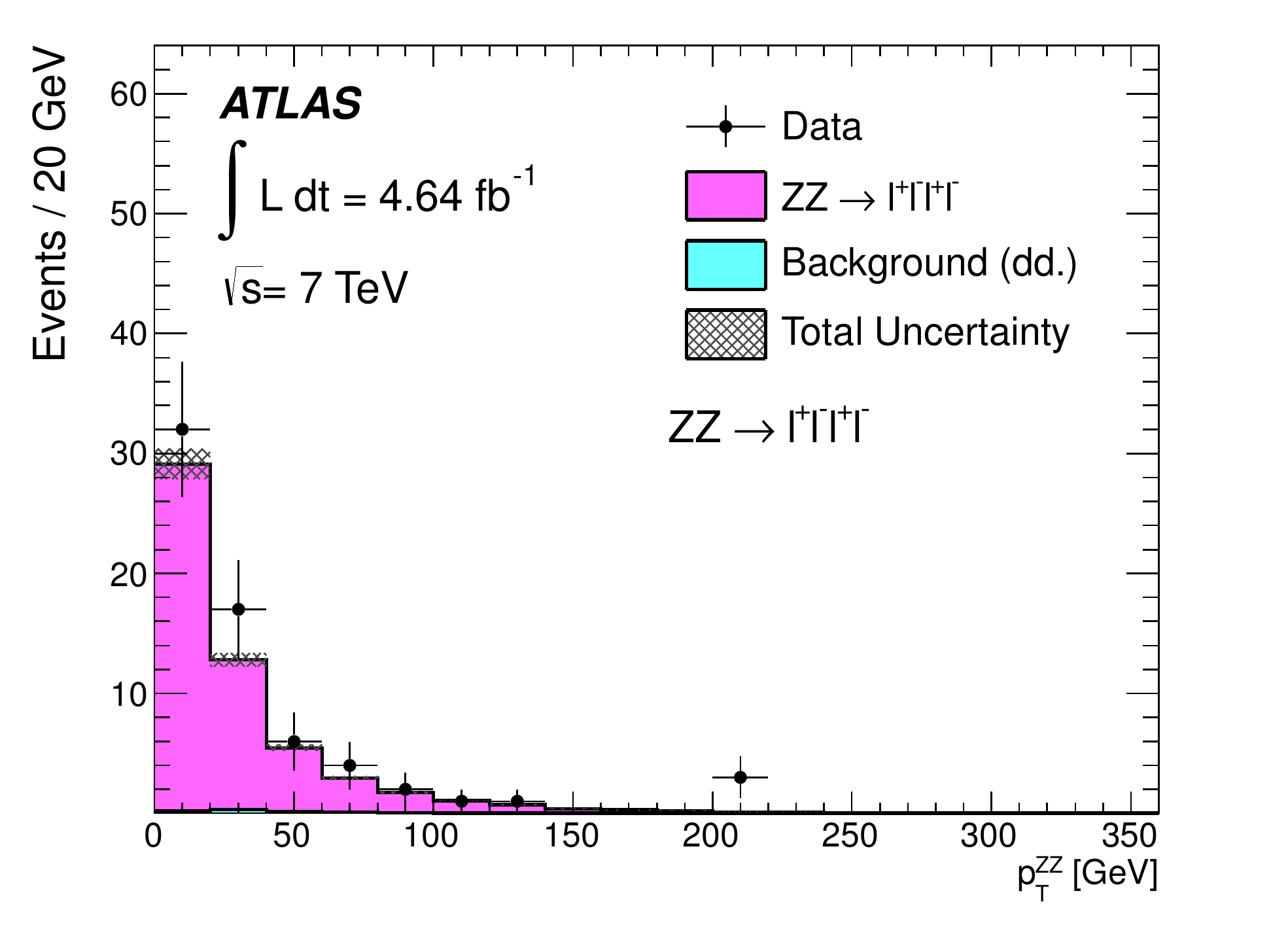}
 }
 \subfigure[]{
 \includegraphics[width=0.47\textwidth]{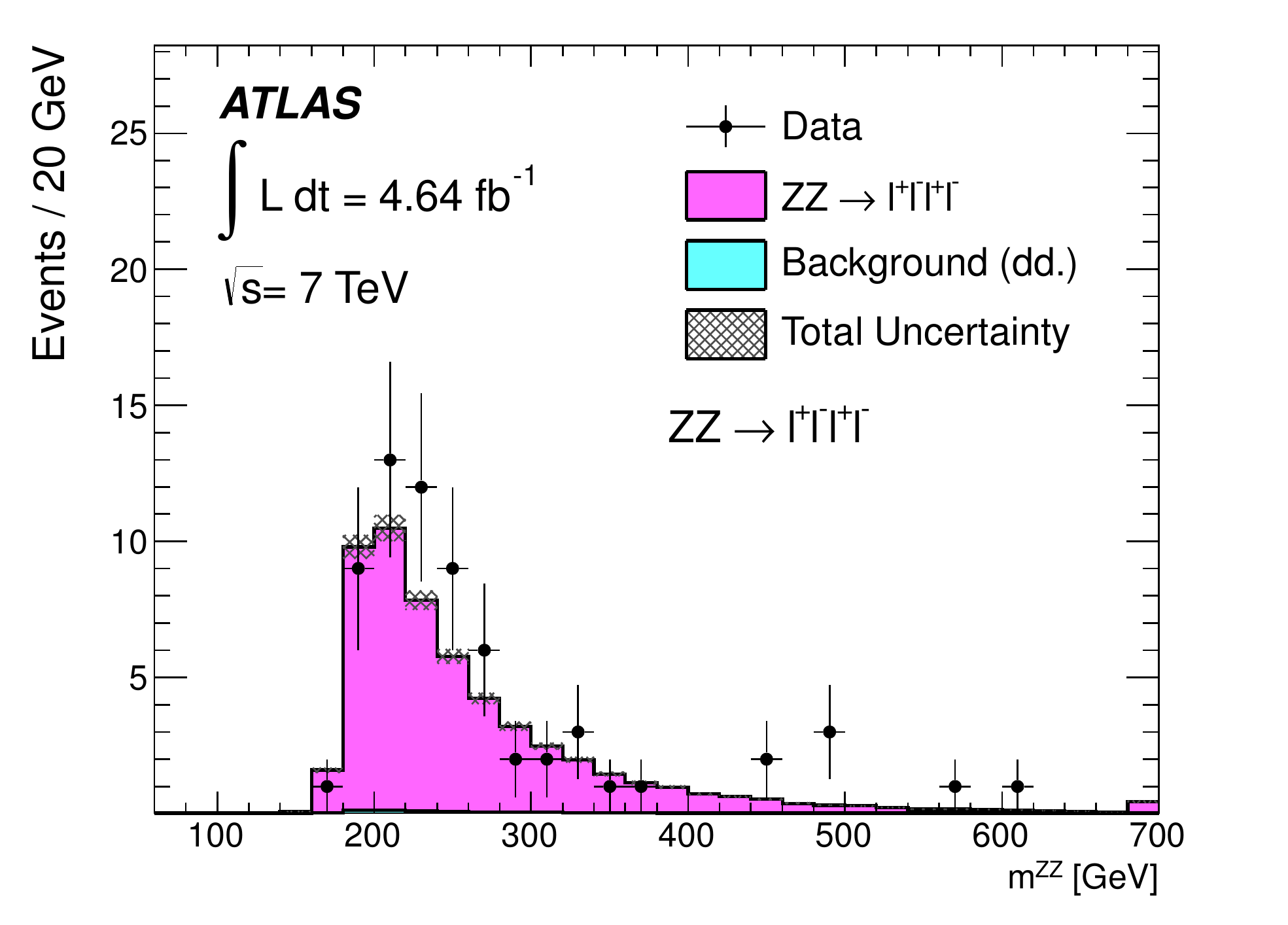}
 }
\caption{\label{fig:kindists_zz}(a) Transverse momentum $\pT^{\ZZ}$ and (b) invariant mass $m^{\ZZ}$ of the 
           four-lepton system for the $ZZ$ selection. The points represent the observed data and the 
           histograms show the prediction from simulation, where the background
           is normalized to the data-driven (dd) estimate as described in section~\ref{sec:Background4l}. The shaded band 
           shows the combined statistical and systematic uncertainty on the prediction. 
}
\end{center}
\end{figure}

\begin{figure}[htbp]
 \begin{center}
 \subfigure[]{
 \includegraphics[width=0.47\textwidth]{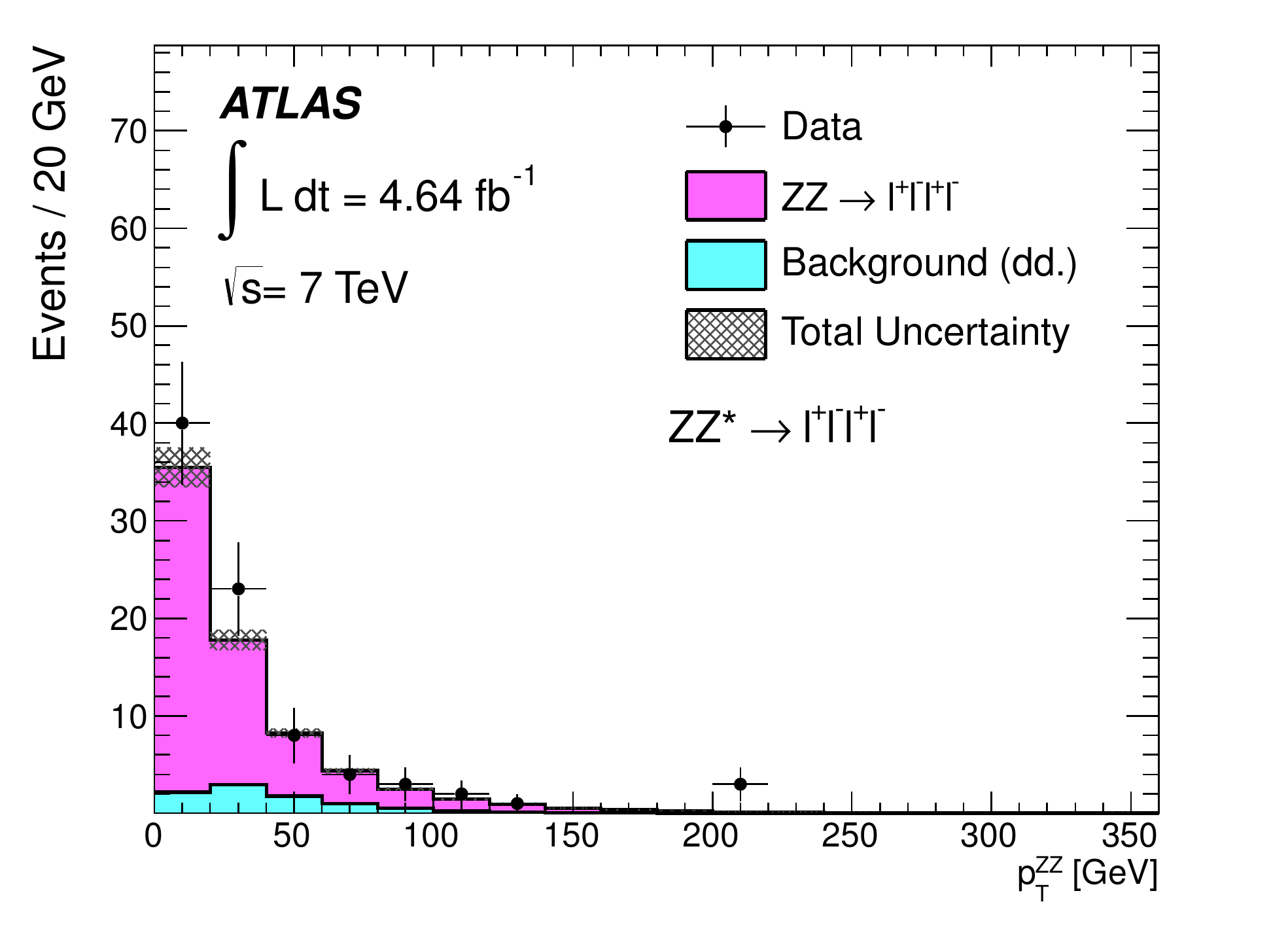}
 }
 \subfigure[]{
 \includegraphics[width=0.47\textwidth]{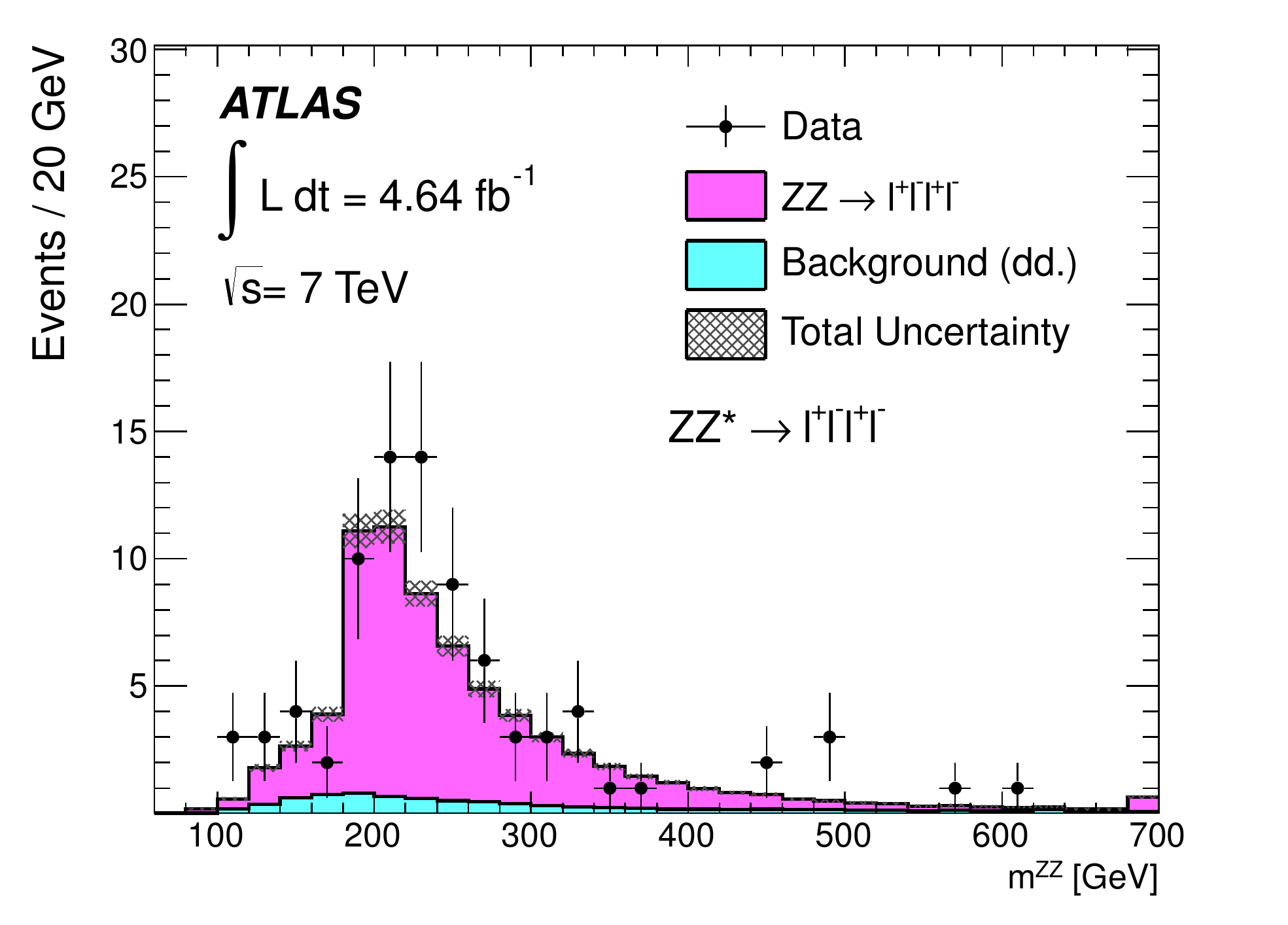}
 }
\caption{\label{fig:kindists_zzs}(a) Transverse momentum $\pT^{\ZZ}$ and (b) invariant mass $m^{\ZZ}$ of the
           four-lepton system for the $ZZ^{*}$ selection. The points represent the observed data and the 
           histograms show the prediction from simulation, where the background
           is normalized to the data-driven (dd) estimate. 
           The shaded band 
           shows the combined statistical and systematic uncertainty on the prediction. 
}
\end{center}
\end{figure}

\begin{figure}[htbp]
\begin{center}
 \subfigure[]{
 \includegraphics[width=0.47\textwidth]{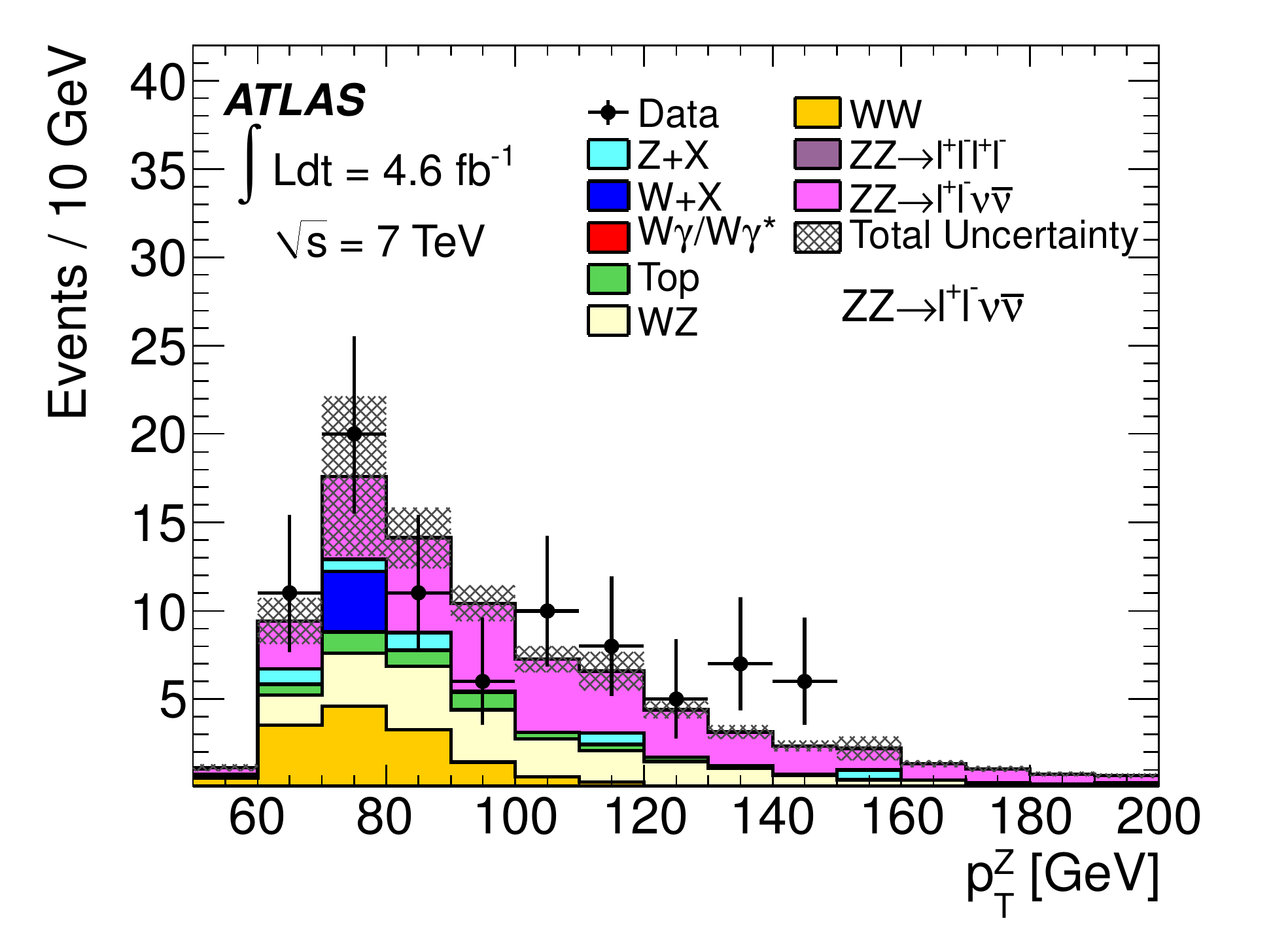}
 }
 \subfigure[]{
 \includegraphics[width=0.47\textwidth]{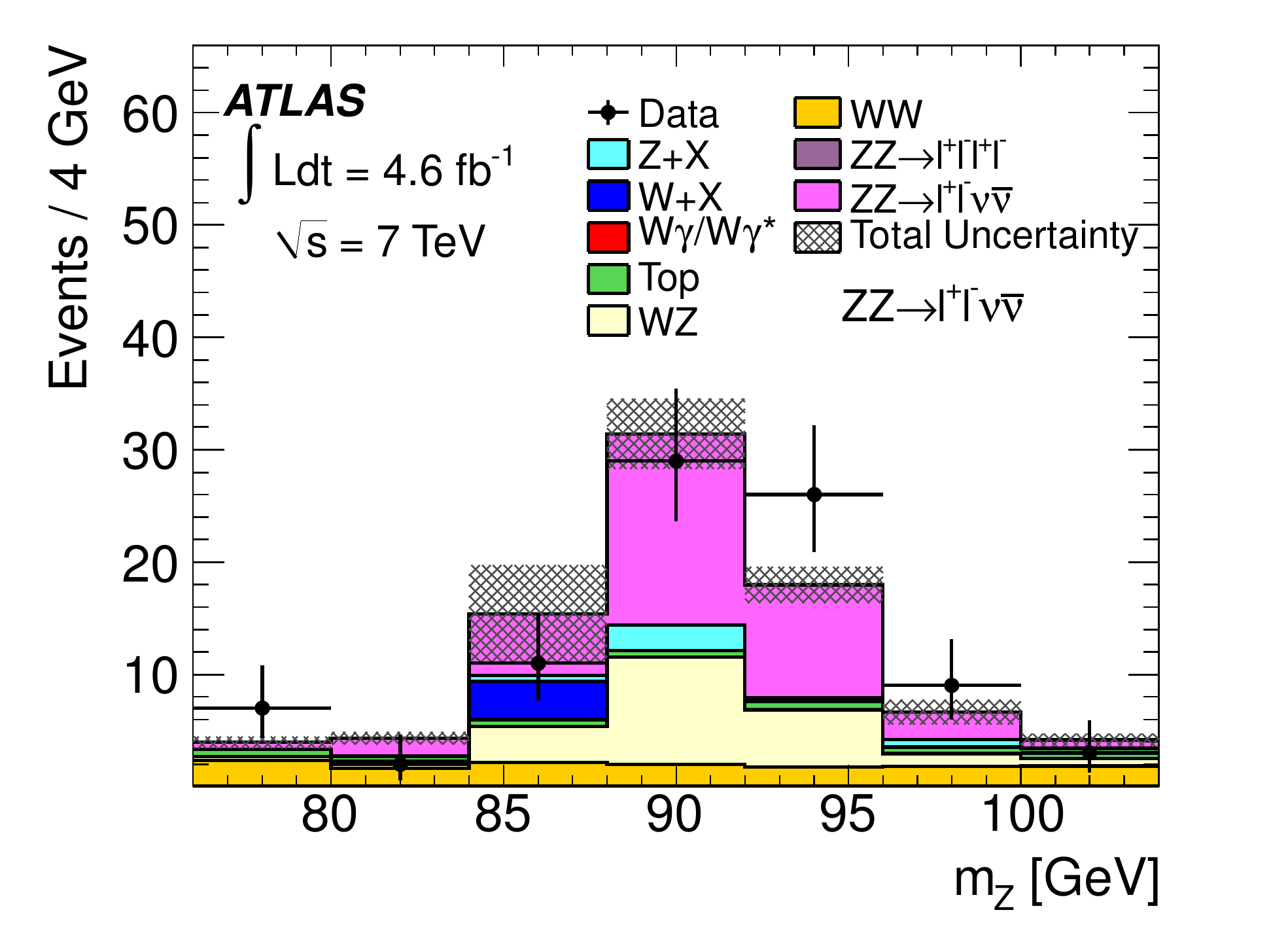}
 }
   \caption{\label{fig:kindists_zz2l2n}(a) Transverse momentum $\pT^{Z}$ and (b) mass $m_{Z}$
           of the two-charged-lepton system for the \zzllvv\ selection. The points represent the observed data and the 
           histograms show the prediction from simulation. The shaded band 
           shows the combined statistical and systematic uncertainty on the prediction. 
   }
 \end{center}
 \end{figure}

The \zzSllll\ and \zzllvv\ fiducial cross sections are determined using a maximum likelihood fitting method,
taking into account the integrated luminosity and the $C_{ZZ}$ correction factors discussed in section~\ref{sec:SignalAcceptance}.
A Poisson probability function is used to model the number of expected events, multiplied by Gaussian 
distribution functions which model the nuisance parameters representing systematic uncertainties. 
The measured fiducial cross sections are: 
\begin{eqnarray*}
\sigma_{ \textrm{\zzllll} }^\mathrm{fid}          &=& \textrm{\fidllll,} \\
\sigma_{ \textrm{\zzsllll} }^\mathrm{fid}        &=& \textrm{\fidsllll,} \\
\sigma_{ZZ\to \ll\nu\bar{\nu}}^\mathrm{fid}  &=& \textrm{\fidllvv}.
\end{eqnarray*}
where \llll\ refers to the sum of the \ee\ee, \ee\mumu\ and \mumu\mumu\ final states and $\ll\nu\bar{\nu}$ refers to the 
sum of the \ee\MET\ and \mumu\MET\ final states\footnote{The \zzllvv\ fiducial region is more restricted compared to the 
\zzSllll\ channel.}.
The expected SM fiducial cross sections, derived from \powhegbox\ and \ggtwozz, are:
\begin{eqnarray*}
\sigma_{ \textrm{\zzllll} }^\mathrm{fid, SM}          &=& \textrm{\theoryfidllll,} \\
\sigma_{ \textrm{\zzsllll} }^\mathrm{fid, SM}        &=& \textrm{\theoryfidsllll,} \\
\sigma_{ZZ\to \ll\nu\bar{\nu}}^\mathrm{fid, SM}  &=& \textrm{\theoryfidllvv}.
\end{eqnarray*}
The measured cross sections are compatible with these theoretical values. 

The total \ZZ\ cross section is calculated by extrapolating to the full phase space while each $Z$ boson is required to have a mass within the \Z\ mass window. 
Both \zzllll\ and \zzllvv\ events are combined in the maximum likelihood fit, 
taking into account the known $Z$ branching fractions~\cite{PDG} and the $A_{ZZ}$ kinematic and geometrical acceptances (section~\ref{sec:SignalAcceptance}). 
Correlated systematic uncertainties between the \zzllll\ and \zzllvv\ channels 
are taken into account in the fit using a single Gaussian for the nuisance parameter for each source of
correlated uncertainty.
The measured value of the total \ZZ cross section is:
\begin{eqnarray*}
\sigma_{ZZ}^\mathrm{tot} &=& \textrm{\totzz.}
\end{eqnarray*}
The result is consistent within errors with the NLO Standard Model total cross section for this
process of \theoryzzmass, where the quoted theoretical uncertainties result from varying the factorization 
and renormalization scales simultaneously by a factor of two and from using the full CT10 PDF error set.

        \subsection{Differential cross sections}\label{sec:Unfolding}
        The differential cross sections present a more detailed comparison of theory to measurement, 
allowing a generic comparison of the kinematic distributions to new theories. 
Variables which are sensitive to new phenomena, such as $p_{\mathrm T}^{Z}$, $m^{ZZ}$ and $\Delta\phi(\ell^{+},\ell^{-})$, 
are used with bin boundaries chosen to maximize sensitivity to nTGCs. At the same time, the bin widths were 
chosen to be commensurate with the resolution. 

The measured 
distributions are unfolded back to the underlying distributions, accounting for the effect of detector resolution, 
efficiency and acceptance, within the fiducial region of each measurement. 
The unfolding procedure is based on a Bayesian iterative algorithm~\cite{bib:iterunfolding}. The algorithm takes as
input a prior for the kinematic distribution and iterates using the posterior distribution as prior for the next iteration. 
The initial prior is taken from the signal Monte Carlo expectation calculated using the 
\powhegbox\ generator and three iterations are performed. The uncertainty on the unfolded distributions is dominated 
by the statistical uncertainty, which is about 30\% in most bins. The systematic uncertainty is no more than 5\% in any bin. 
The dependence of the unfolded cross sections on the choice of the initial prior is tested by unfolding the measured 
distributions using a different generator (\sherpa). 
The difference between the two is taken as a systematic uncertainty to account for differences in generator modelling 
(e.g. QCD radiation). 
The difference in unfolded distributions between three iterations and four iterations is much lower than the statistical 
uncertainty and it is taken as a further uncertainty on the unfolding procedure. 
Systematic uncertainties related to detector effects (e.g. lepton reconstruction efficiency) are evaluated using 
pseudo-experiments.

Figures~\ref{fig:unfolded} to~\ref{fig:unfoldedM4l} show the differential cross sections normalized to the fiducial 
cross sections for the $p_{\mathrm T}^{Z}$ and $\Delta\phi(\ell^+,\ell^-)$ of the leading $Z$ boson, and for 
the mass (transverse mass) of the \ZZ\ system for the 
\zzllll\ (\zzllvv) selection. 
 The Standard Model prediction is consistent with the measurement in each case.

\begin{figure}[htbp]
\begin{center}
\subfigure[]{\includegraphics[width=0.48\textwidth]{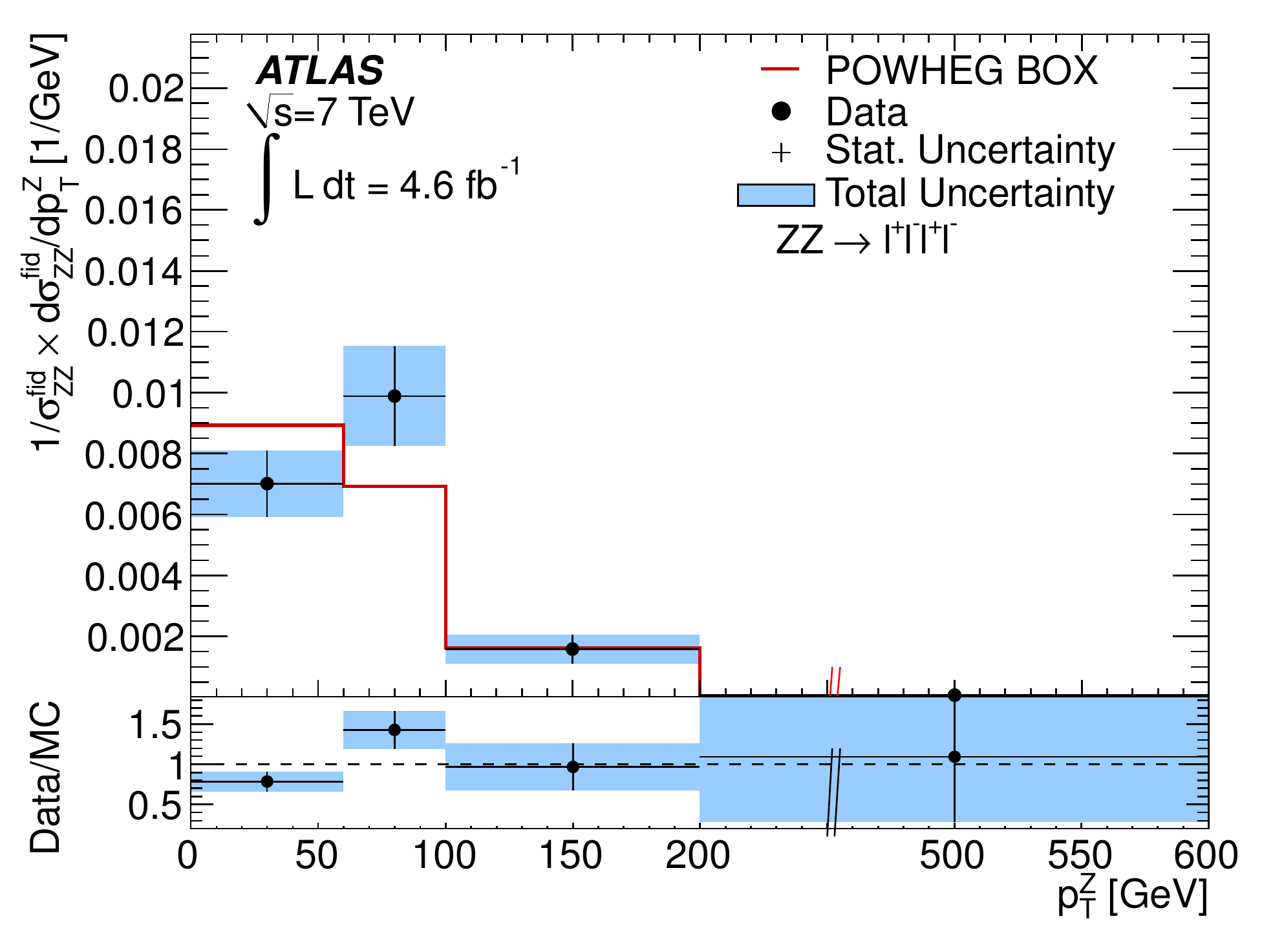}}
\subfigure[]{\includegraphics[width=0.48\textwidth]{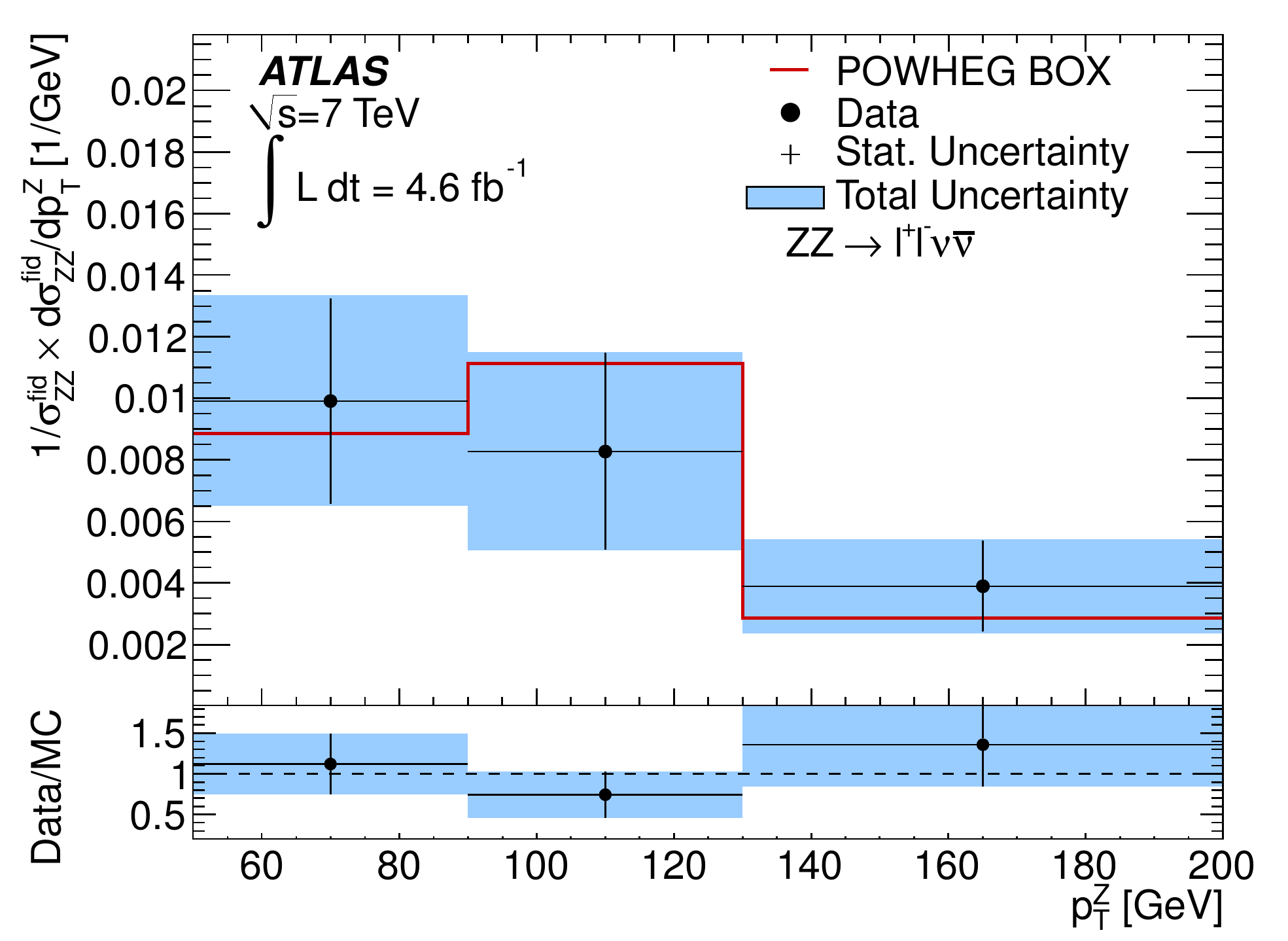}}
\caption{\label{fig:unfolded}Unfolded $ZZ$ fiducial cross sections in bins of the
\pT\ of the leading $Z$ boson for (a) the \zzllll\ selection, where a discontinuity is indicated by the 
parallel pairs of lines, and (b) the \zzllvv\ selection.}
\end{center}
\end{figure}

\begin{figure}[htbp]
\begin{center}
\subfigure[]{\includegraphics[width=0.48\textwidth]{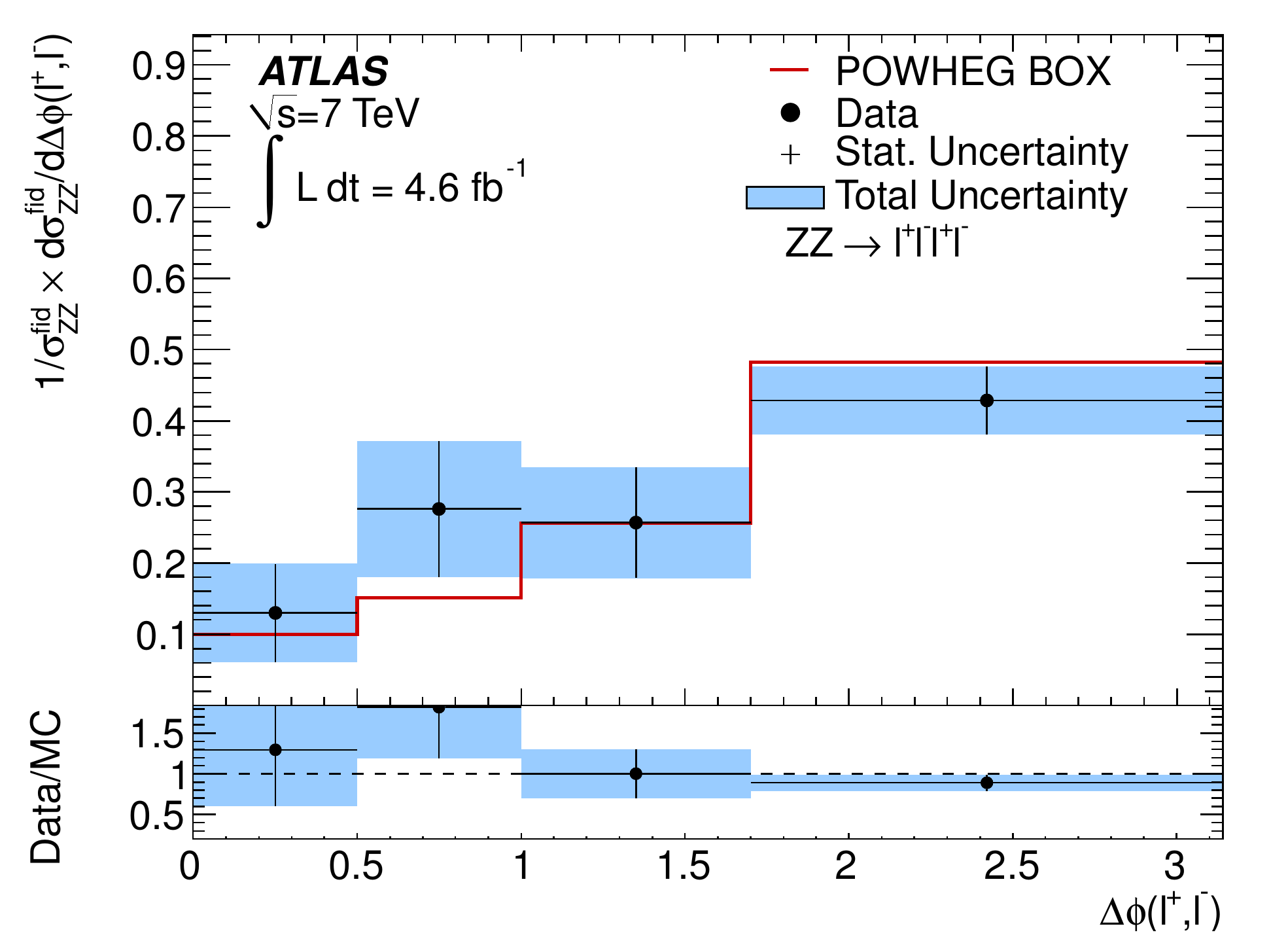}}
\subfigure[]{\includegraphics[width=0.48\textwidth]{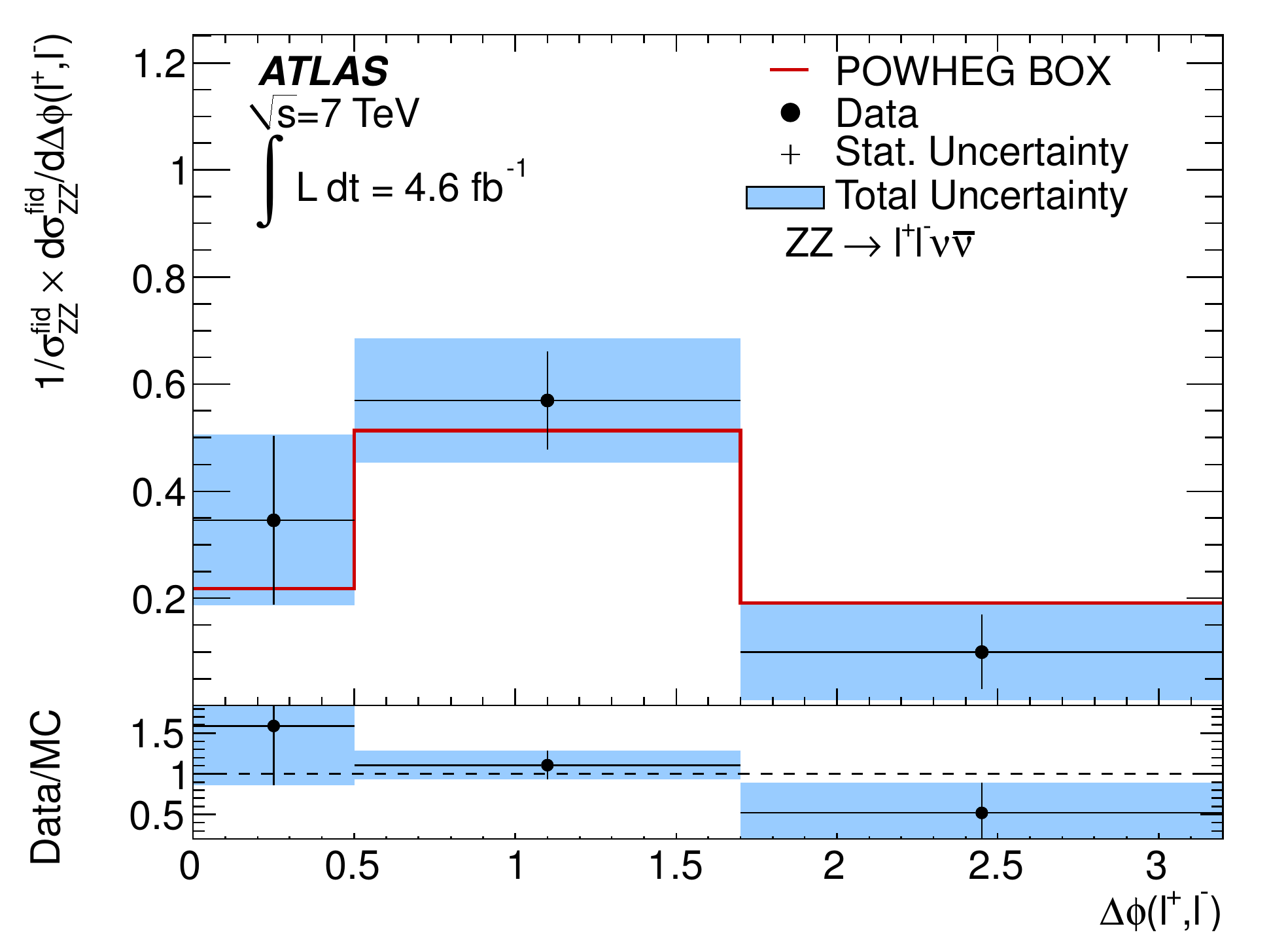}}
\caption{\label{fig:unfoldedDphi}Unfolded $ZZ$ fiducial cross sections in bins of the $\Delta\phi(\ell^+,\ell^- )$ of the leading $Z$ boson for (a) the \zzllll\ selection and (b) the \zzllvv\ selection.}
\end{center}
\end{figure}

\begin{figure}[htbp]
\begin{center}
\subfigure[]{\includegraphics[width=0.48\textwidth]{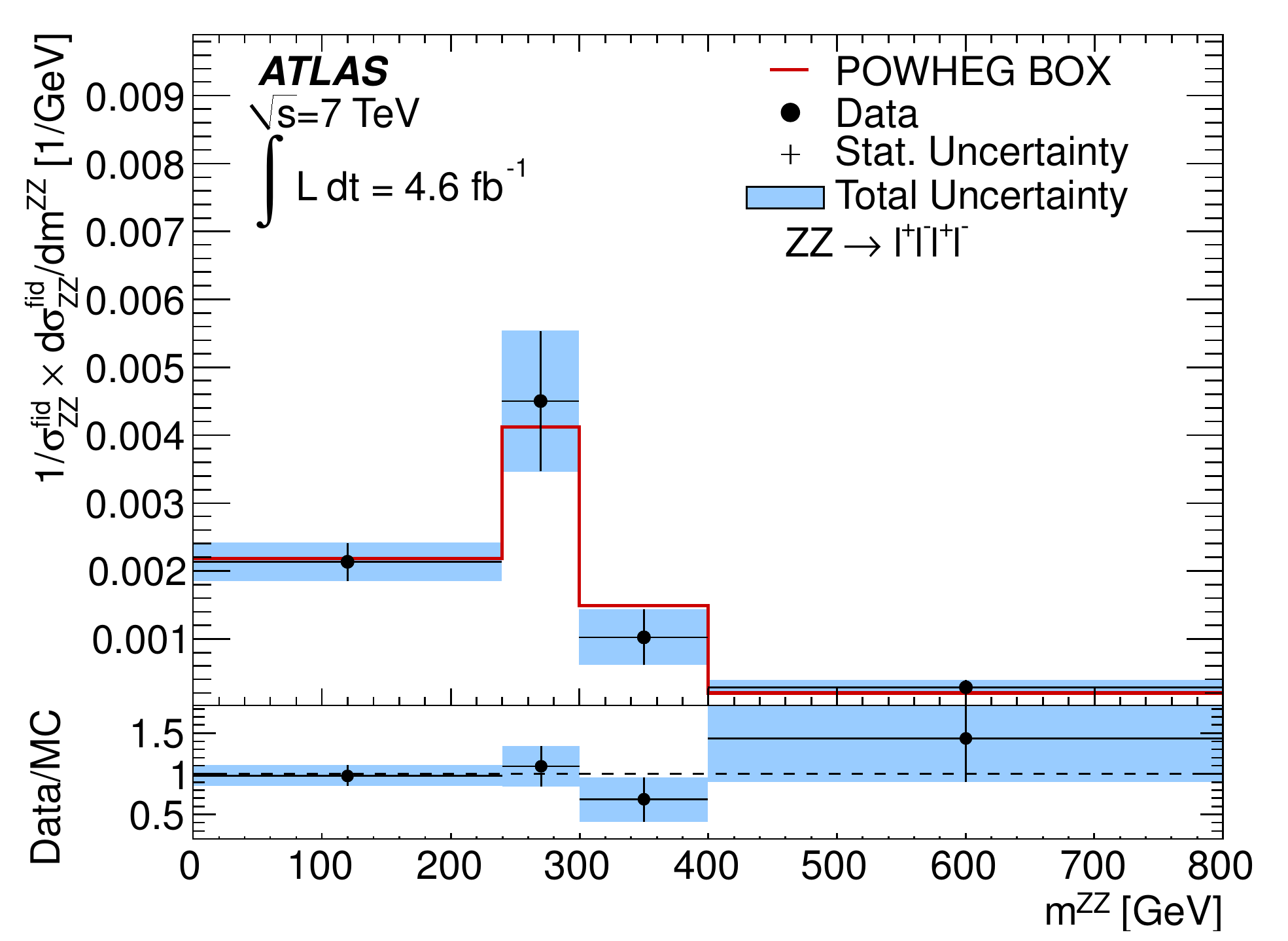}}
\subfigure[]{\includegraphics[width=0.48\textwidth]{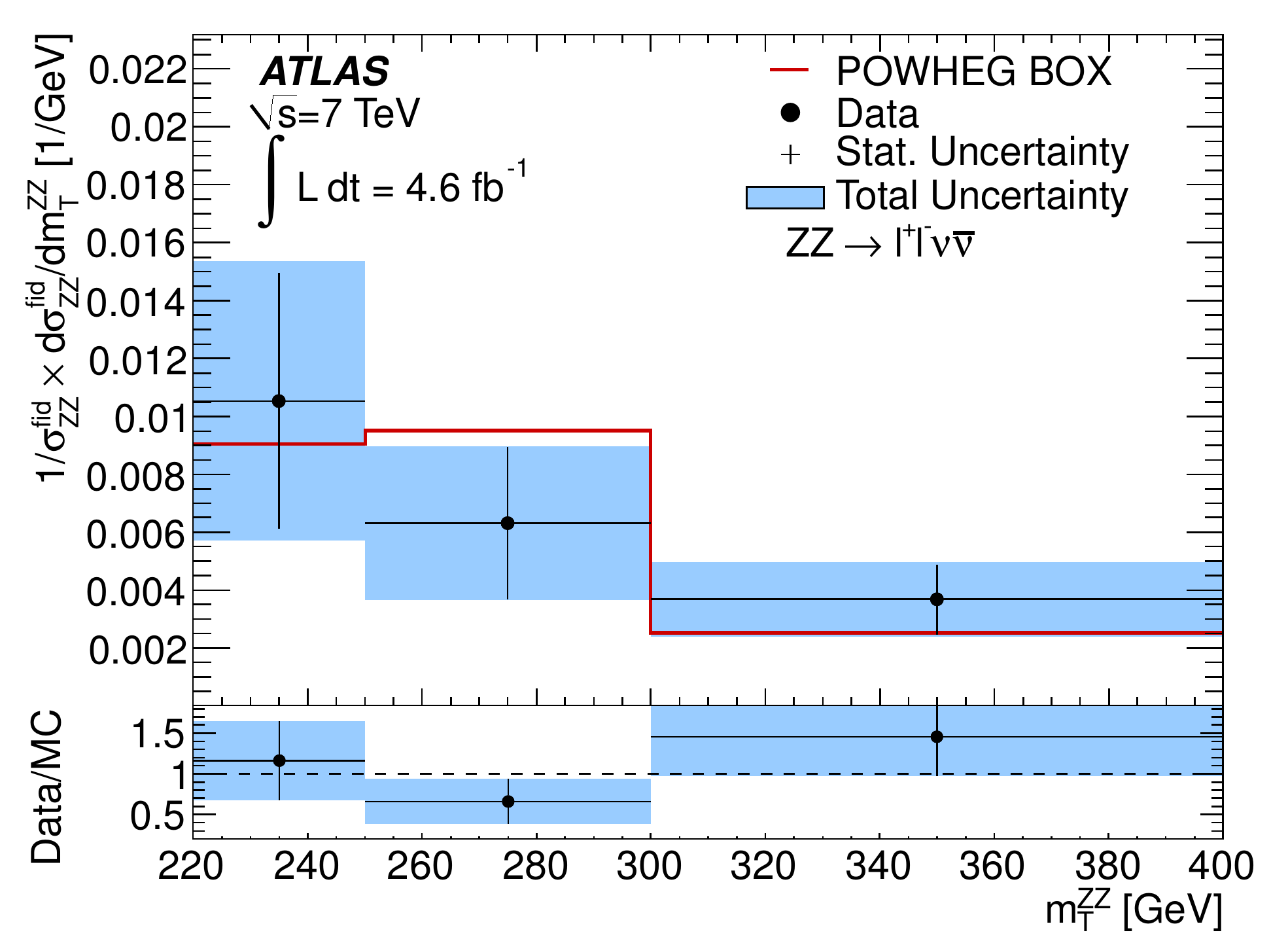}}
\caption{\label{fig:unfoldedM4l}Unfolded $ZZ$ fiducial cross sections in bins of 
(a) $m^{ZZ}$ for the \zzllll\ selection and (b) $m^{ZZ}_{\rm T}$ for the \zzllvv\ selection.}
\end{center}
\end{figure}

        \subsection{Anomalous neutral triple gauge couplings}\label{sec:TGC}
        
Anomalous nTGCs for on-shell \ZZ\ production can be parameterized by two 
CP-violating ($f_4^V$) and two CP-conserving ($f_5^V$) complex parameters 
(where $V = \Z, \gamma$) which are zero in the Standard Model~\cite{Baur:2000ae}. 
A form-factor parameterization is introduced leading to couplings which vanish at 
high parton centre-of-mass energy $\sqrt{\hat s}$: $f_i^V = f_{i0}^{V}/(1+\hat s/\Lambda^{2})^{n}$, 
ensuring partial-wave unitarity. 
Here, $\Lambda$ is the energy scale at which physics beyond the Standard Model would be directly observable, $f^{V}_{i0}$ are the
low-energy approximations of the couplings, and $n$ is the form-factor
power.
Values of $n = 3$ and $\Lambda = 3 \TeV$ are chosen, so that expected limits
are within the values allowed by requiring that unitarity is not violated at LHC energies~\cite{Baur:2000ae}.
The results with an energy cutoff $\Lambda=\infty$ (i.e. without a form factor) are also presented as a 
comparison in the unitarity violating scheme.

Limits on anomalous nTGCs are determined using the observed and expected numbers 
of \zzllll\ and \zzllvv\ events binned\footnote{The raw 
(i.e. not unfolded) differential event yields are used, to avoid 
introducing theory dependence.} in $p_{\rm{T}}^{Z}$, 
as seen in table~\ref{tab:eventYield_perPtZbin}.
Figure~\ref{fig:pt_tgc} shows the observed $p_{\rm{T}}^{Z}$ distributions, together with the SM expectation 
and the predicted distributions for nTGC values close to the previous limits obtained by ATLAS~\cite{ATLAS_ZZ4l:1fb2011}. 
Using an increased data sample compared with our previous measurement, including the \zzllvv\ channel, and exploiting the differential event yields,  
the precision is expected to improve by about a factor of five. 
The dependency of the couplings on the expected number of events in each $p_{\rm{T}}^{Z}$ bin is parameterized using
fully simulated events, generated with  \sherpa~\cite{bib:sherpa},  subsequently reweighted using the Baur--Rainwater~\cite{Baur:2000ae, Baur:1994au} and BHO~\cite{bib:bho} MC generators.
The next-to-leading-order matrix elements with their nTGC dependence have been extracted 
from the BHO MC generator for $2\ra5$ events and the Baur--Rainwater MC generator for $2\ra4$ 
events and introduced into a framework~\cite{Bella:2008wc} that enables 
a calculation of the amplitude given the four vectors and the identity of the incoming and outgoing particles from the hard process. 

\begin{table}
\begin{center}
\begin{tabular}{c c c c}
\hline
    & Expected background & Expected \ZZ\ signal & Observed events \\
\hline
\zzllll \\
\hline
$0<\pt^Z <60$ GeV    & $0.6    \pm0.8   \pm0.5  $   & 27.9 $\pm$ 0.2 $\pm$ 2.0 & 28 \\
$60<\pt^Z <100$ GeV  & $0.2   \pm0.2   \pm0.2  $   & 14.6 $\pm$ 0.2 $\pm$ 1.2 &  25 \\
$100<\pt^Z <200$ GeV & $0.1   \pm0.1   \pm0.1  $   & 9.3 $\pm$ 0.1 $\pm$ 0.9 &  11 \\
$\pt^Z > 200 $ GeV     & $0.01  \pm0.01  \pm0.01 $   & 1.6 $\pm$ 0.1 $\pm$ 0.3 &  2 \\
\hline 
\zzllvv           \\ 
\hline
$50<\pt^Z<90$ GeV  & $26.0\pm4.5\pm1.1$ & $13.6\pm0.2\pm1.3$ & $42$ \\
$90<\pt^Z<130$ GeV & $16.0\pm2.8\pm0.7$ & $15.7\pm0.3\pm1.7$ & $29$ \\
$\pt^Z > 130$ GeV   & $4.9\pm1.8\pm0.2$ & $10.1\pm0.1\pm1.5$ & $16$ \\
\hline
\end{tabular}
\caption{Total background, expected signal and observed events
as a function of the \pt\ of the leading \Z\ for the \zzllll\ and \zzllvv\ selections.
For the expected signal and background events, the first uncertainty is statistical and the second is systematic.}
\label{tab:eventYield_perPtZbin}
\end{center}
\end{table}

\begin{figure}[htbp]
\begin{center}
\subfigure[] {
\includegraphics[width=0.47\textwidth]{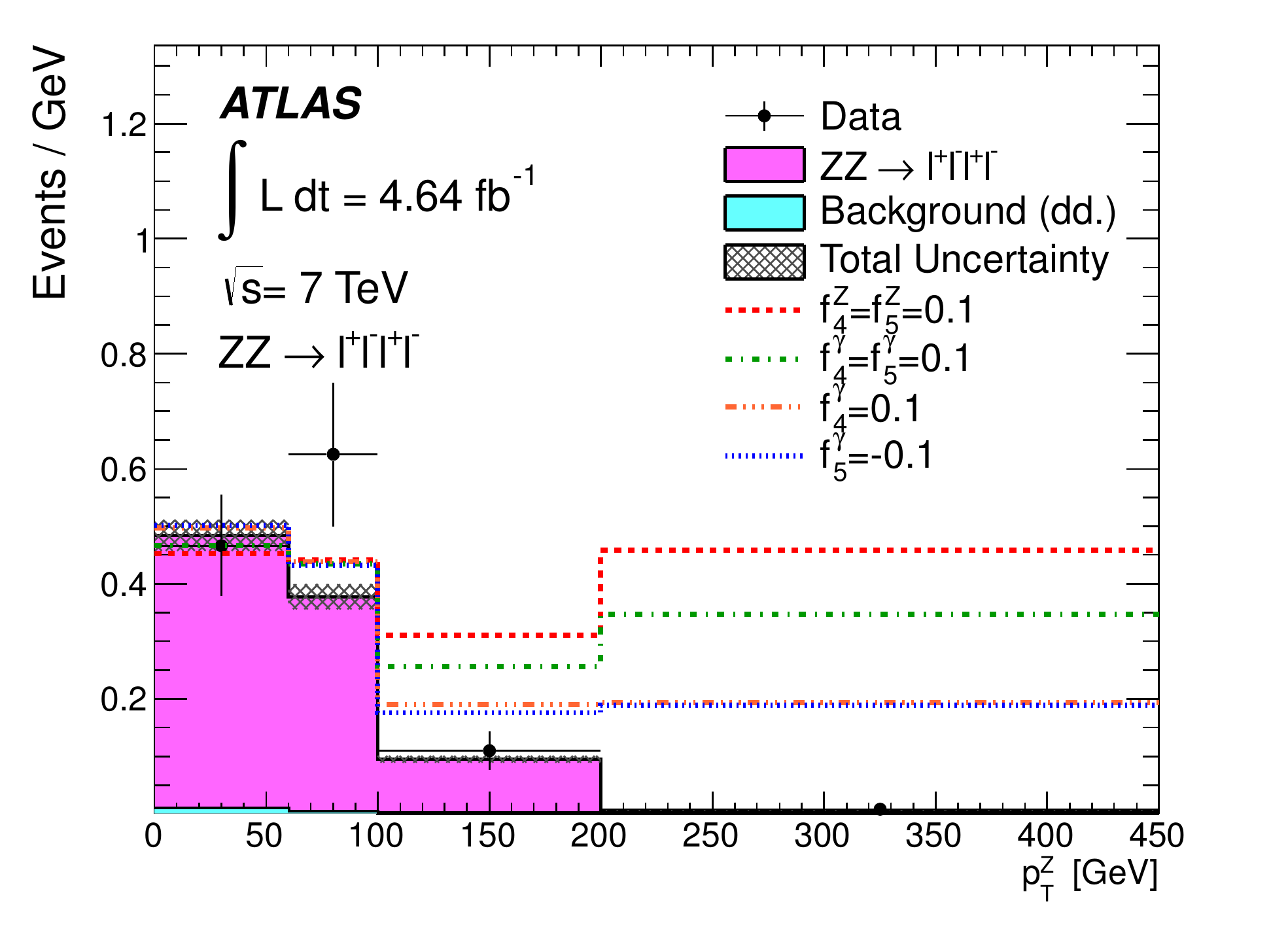}
}
\subfigure[] {
\includegraphics[width=0.47\textwidth]{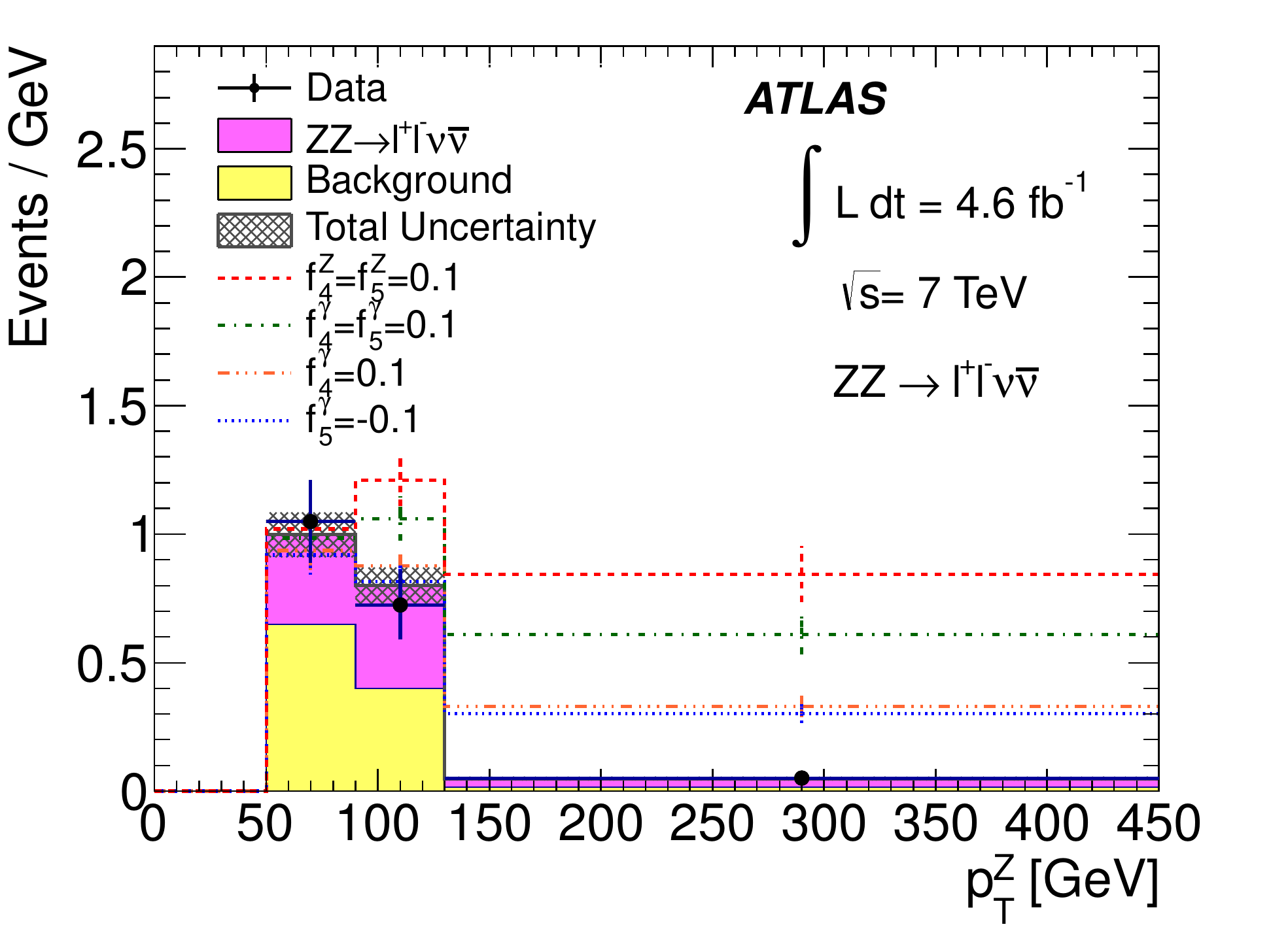}
}
\caption{\label{fig:pt_tgc} The leading $Z$ boson transverse momentum distributions for 
(a) the \zzllll\ selection and (b) the \zzllvv\ selection.
The observed distributions are shown as filled circles, 
the SM expected signal and background are shown as filled histograms, and the predicted 
distributions for four different nTGC samples with form factor 
 scales of $\Lambda=3$ TeV and nTGC coupling values
 set near the edge of the exclusion set in the 1 fb$^{-1}$ analysis~\cite{ATLAS_ZZ4l:1fb2011} are shown as dashed lines.
}
\end{center}
\end{figure}

Confidence intervals for the anomalous triple gauge couplings are determined using the maximum profile
likelihood ratio. Limits are set on each coupling, assuming all of the other
couplings are zero (as in the Standard Model), and on pairs of couplings assuming the
remaining two couplings are zero. 
The profile likelihood ratio is calculated for the data, and also for 10000 pseudo-experiments 
generated using the expected number of events at each point in the one- or two-dimensional nTGC parameter space. 
A point is rejected if more than 95\% of the pseudo-experiments have a larger profile likelihood ratio value than the 
one observed in data. The systematic errors are included as nuisance parameters. 

\begin{table}
\centering
  \begin{tabular}{lcccc}  
    \hline    \hline
     $\Lambda$  & $f_{40}^{\gamma}$ & $f_{40}^{Z}$ & $f_{50}^{\gamma}$ & $f_{50}^{Z}$\\\hline  
     $3$ \TeV  & $[-0.022, 0.023]$  & $[-0.019, 0.019]$ & $[-0.023,0.023]$ & $[-0.020, 0.019]$ \\  
     $\infty$  & $[-0.015, 0.015]$  & $[-0.013, 0.013]$ & $[-0.016,0.015]$ & $[-0.013, 0.013]$ \\  
    \hline    \hline
  \end{tabular}

  \caption{\label{ta:TGCLimits}
           One-dimensional 95\%\ confidence intervals for anomalous neutral gauge boson couplings, where 
           the limit for each coupling assumes the other couplings
           are fixed at zero, their SM value.  
           Limits are presented for form factor scales of $\Lambda = 3$ \TeV\ and $\Lambda = \infty$ and include 
           both statistical and systematic uncertainties; the statistical uncertainties are dominant.
          }

\end{table}

\begin{figure}[htbp]
\begin{center}
\subfigure[]{
\includegraphics[width=0.39\textwidth]{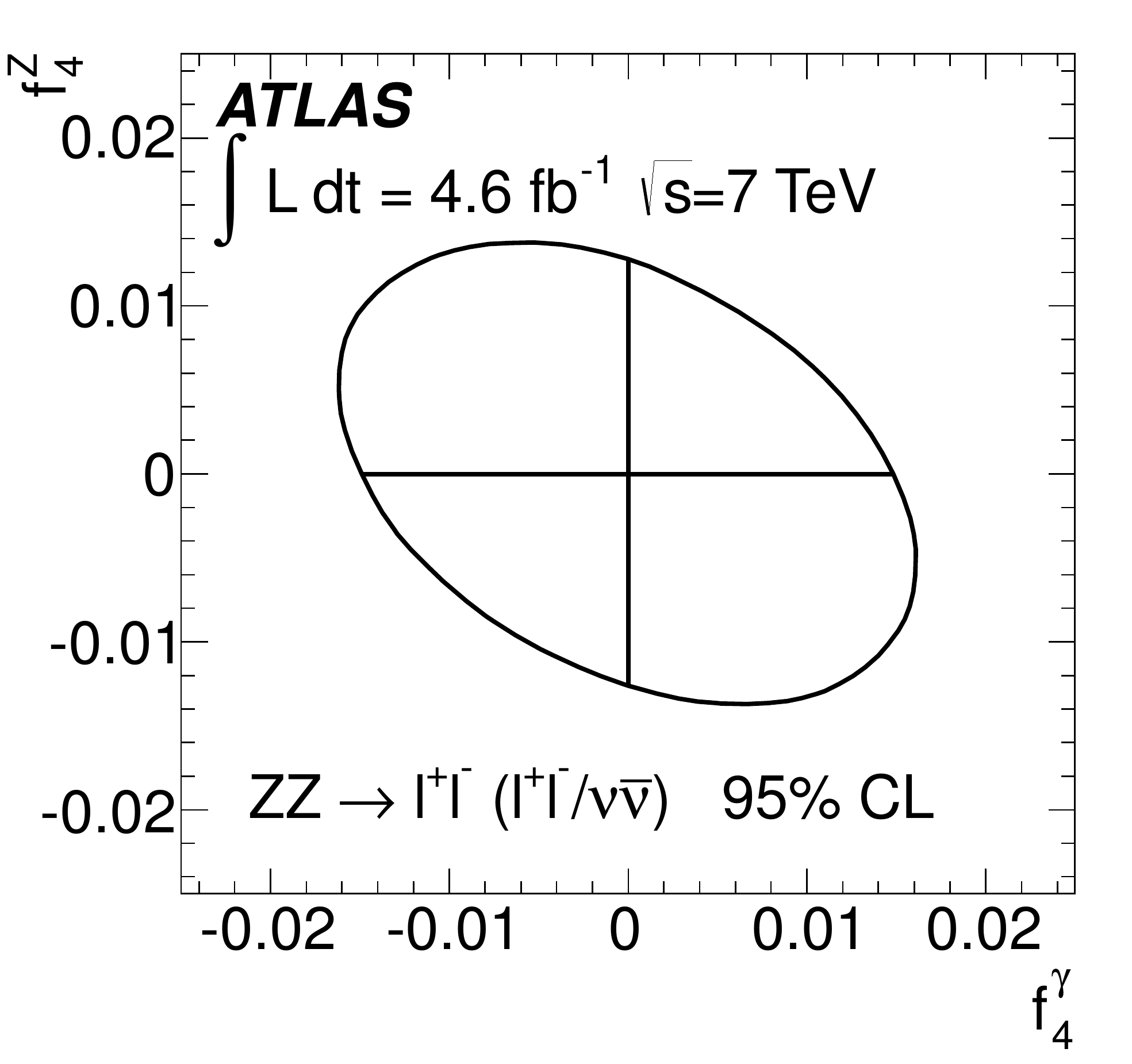}
}
\subfigure[]{
\includegraphics[width=0.39\textwidth]{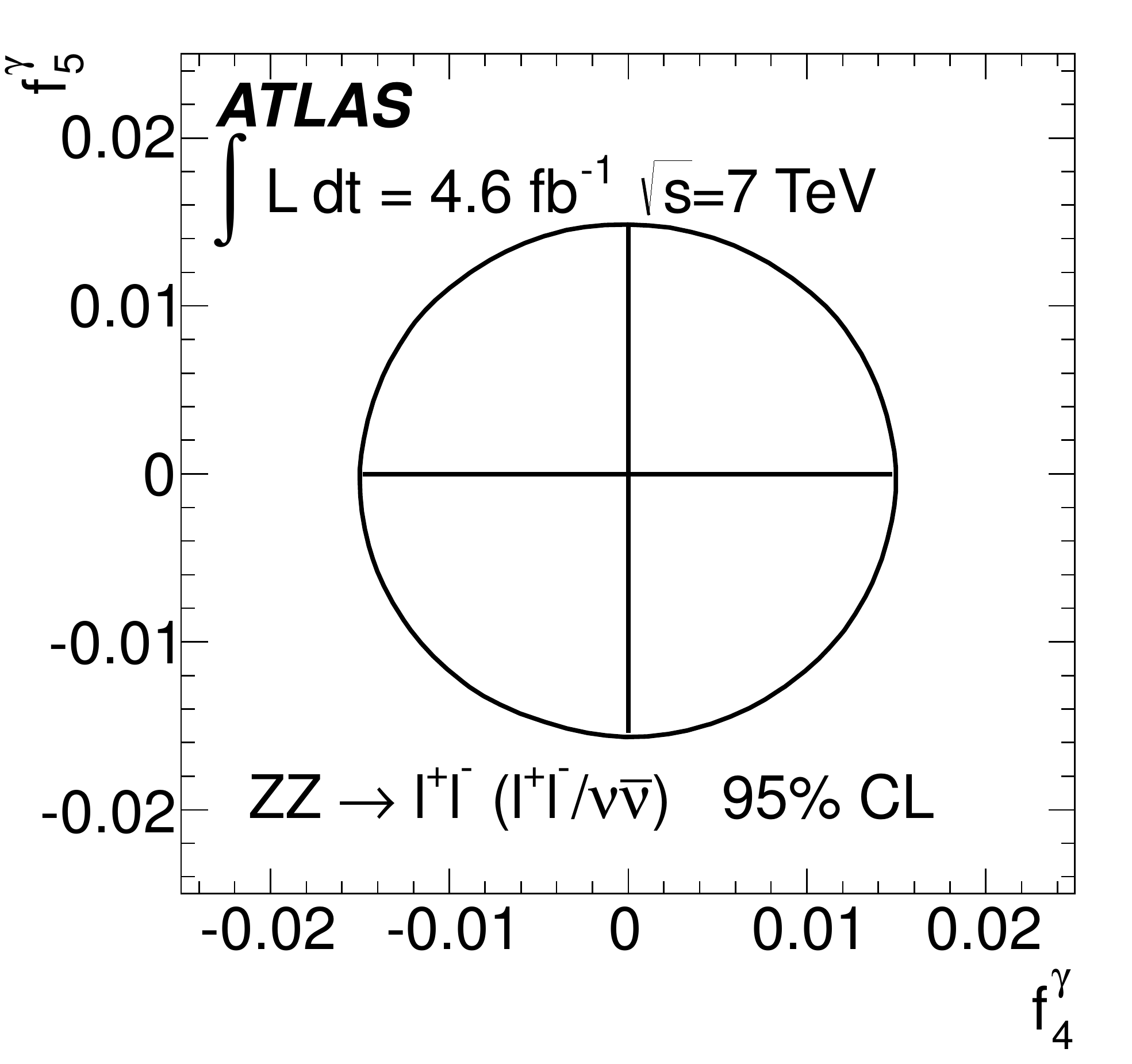}
}
\subfigure[]{
\includegraphics[width=0.39\textwidth]{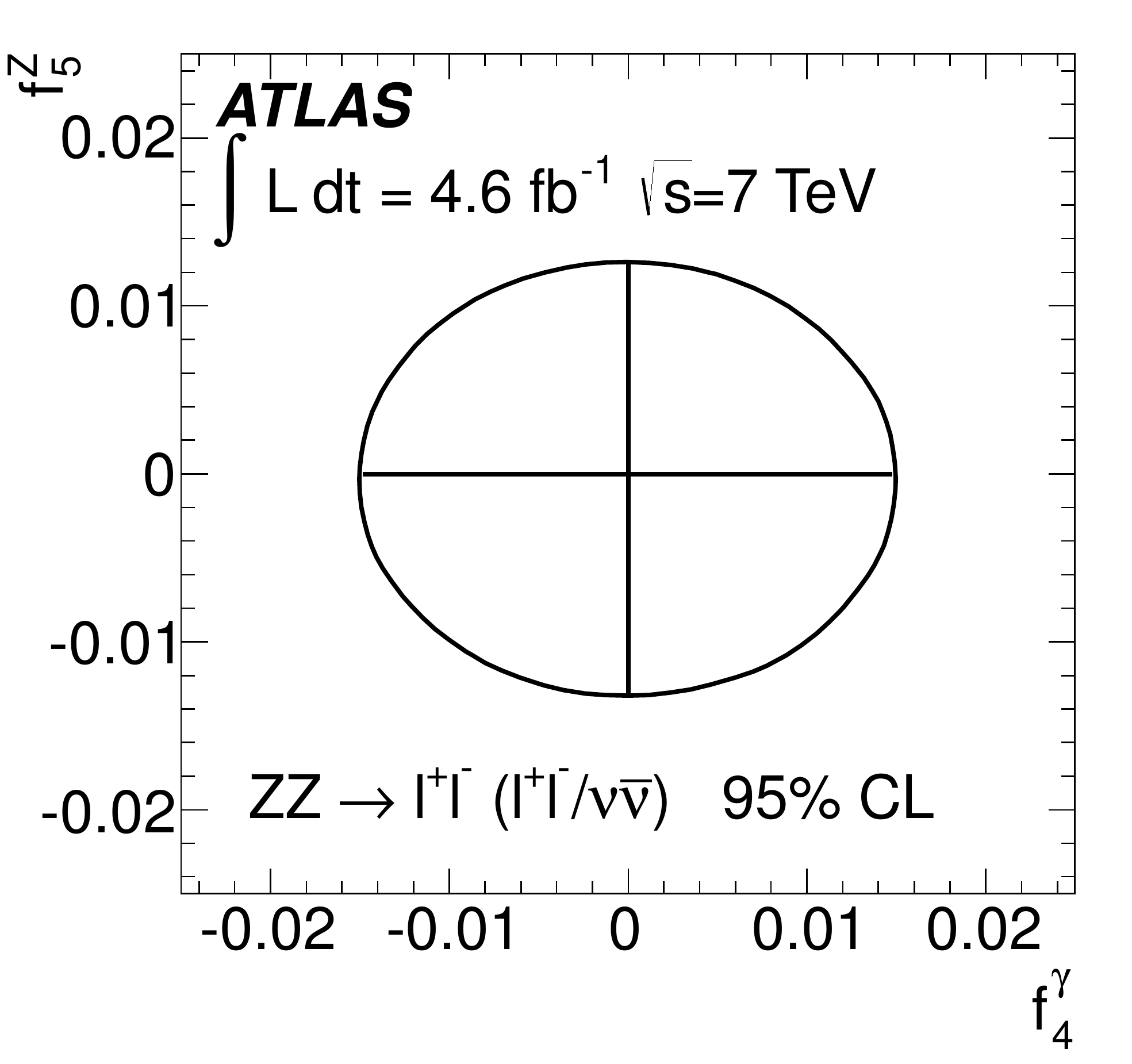}
}
\subfigure[]{
\includegraphics[width=0.39\textwidth]{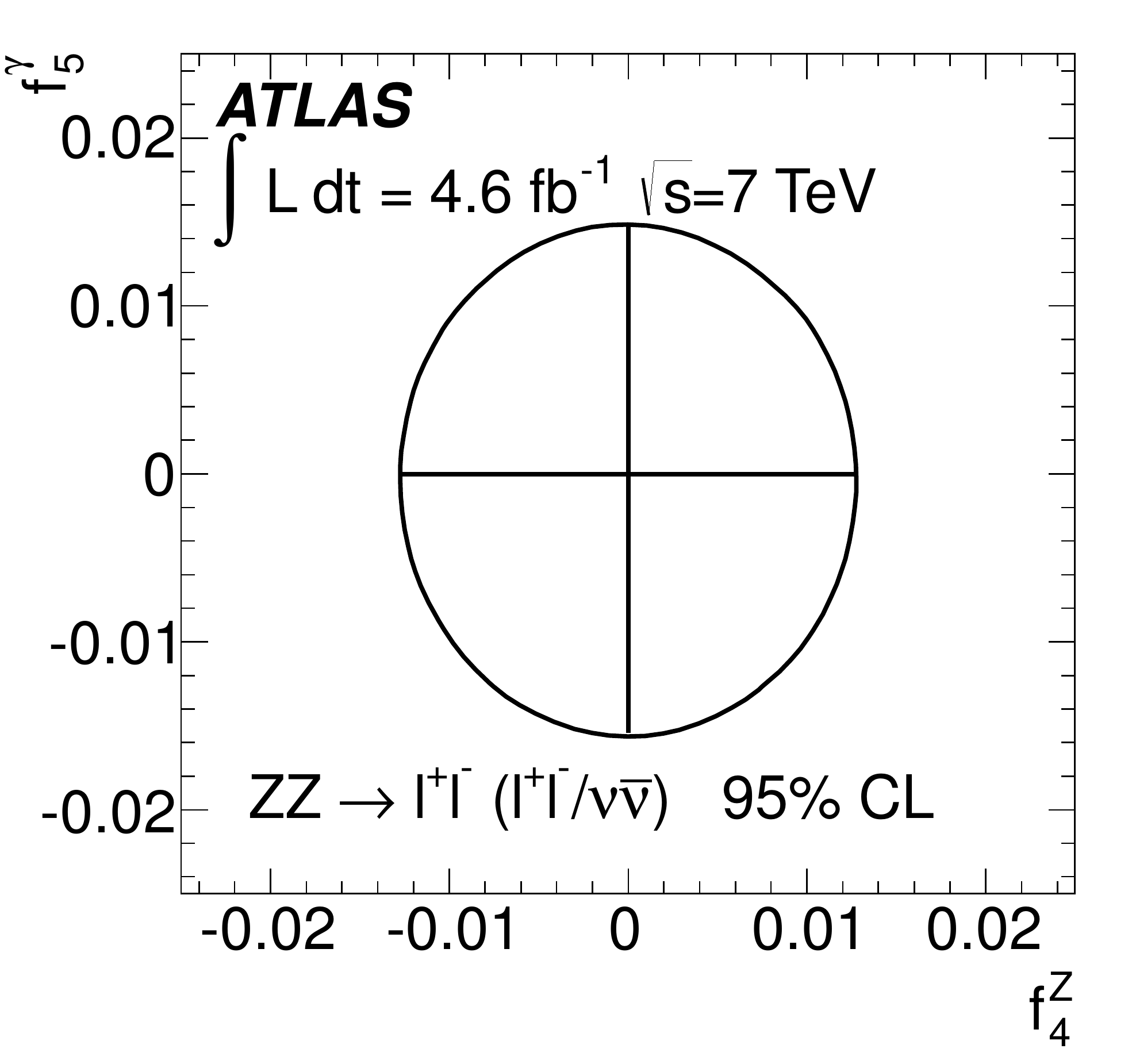}
}
\subfigure[]{
\includegraphics[width=0.39\textwidth]{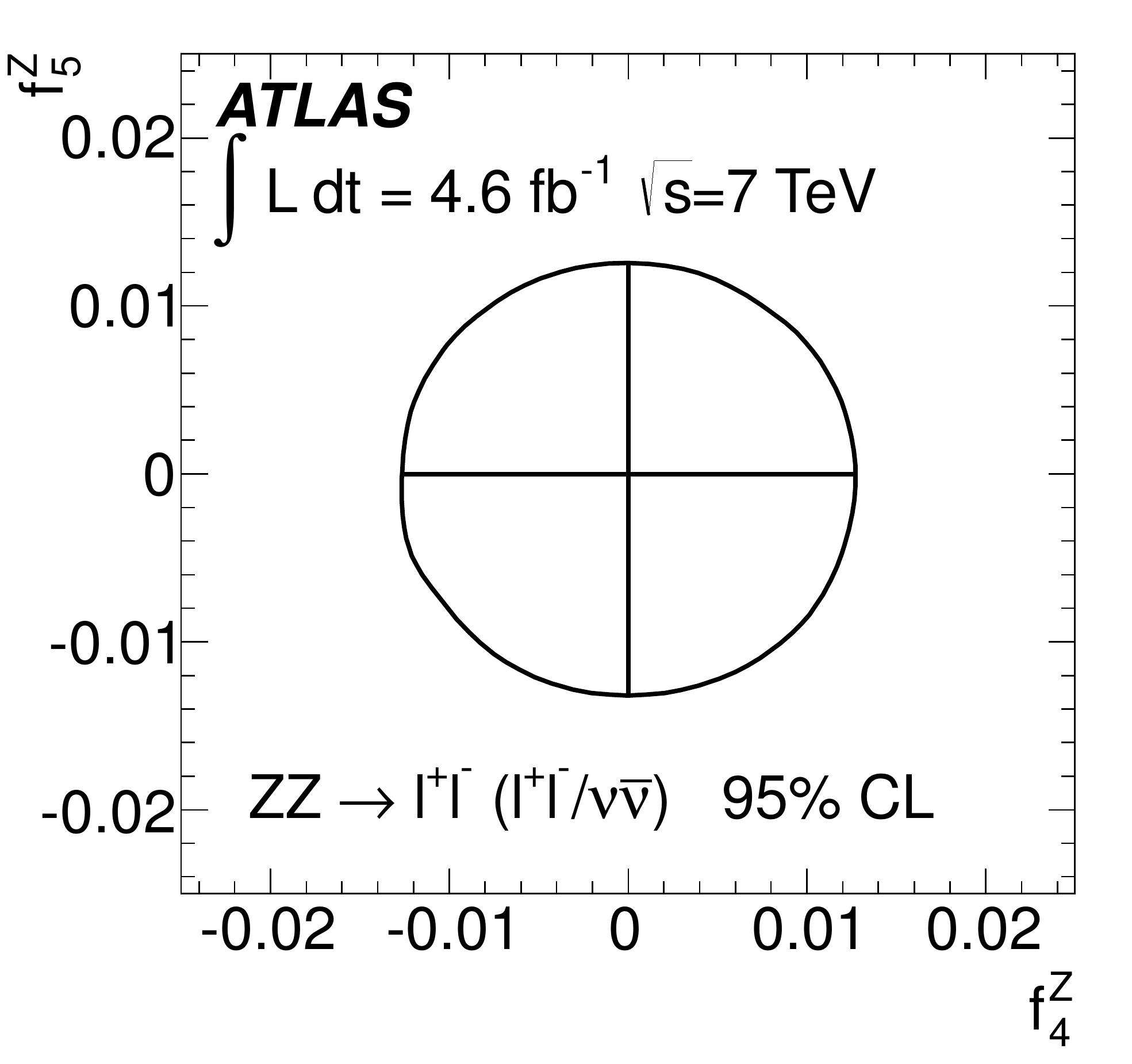}
}
\subfigure[]{
\includegraphics[width=0.39\textwidth]{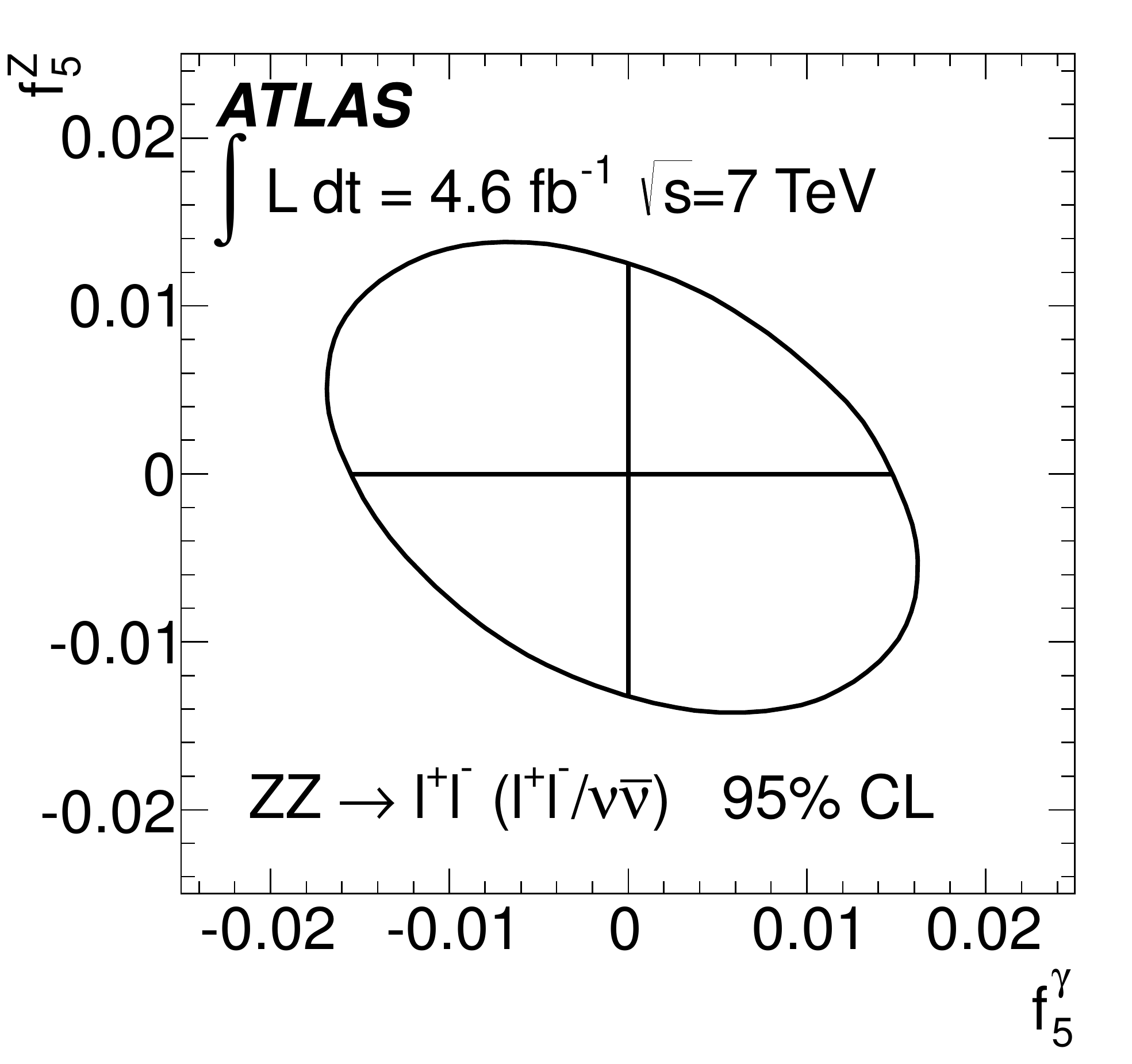}
}
\caption{\label{fig:tgc_limits}
         Two-dimensional triple gauge coupling limits for form factor scale $\Lambda=\infty$. The one-dimensional triple gauge coupling limits are 
         shown as vertical and horizontal lines inside the two-dimensional ellipses, whose shape is determined by the theoretical correlations.  
         For each two-dimensional limit the other TGC parameters are assumed to be zero. 
         Since most of the sensitivity of the measurement is contained in a single bin, the likelihood ratio used to obtain the two-dimensional limits has
         one effective degree of freedom.
}
\end{center}
\end{figure}

The resulting limits for each coupling are listed in table~\ref{ta:TGCLimits}.
Two-dimensional 95\% confidence intervals\footnote{Since most of the sensitivity of the measurement is 
contained in a single bin, the likelihood ratio used to obtain the two-dimensional limits has 
one effective degree of freedom.} are shown in figure~\ref{fig:tgc_limits}.
The one-dimensional limits are more stringent than those derived from measurements at
LEP~\cite{bib:LEPEW2006} and the Tevatron~\cite{bib:D0_ZZ1} and previously by ATLAS~\cite{ATLAS_ZZ4l:1fb2011};
it should be noted that
the limits from LEP do not use a form factor, and those from the Tevatron use
$\Lambda = 1.2\TeV$. A comparison of the LHC limits with those 
derived from LEP and Tevatron is shown in figure~\ref{fig:tgccomb}.

 \begin{figure}[htbp]
 \begin{center}
  \includegraphics[width=0.8\textwidth]{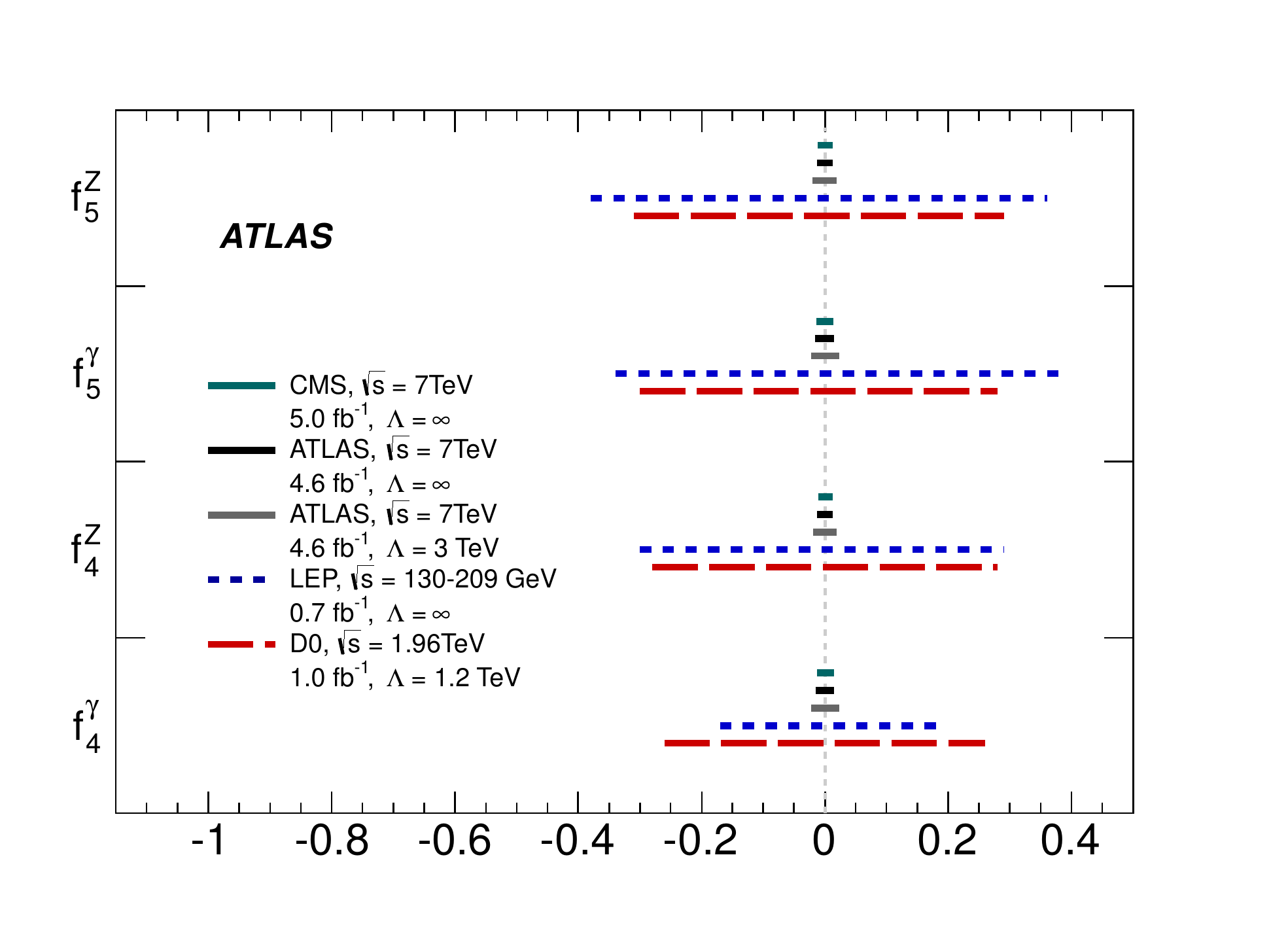}\hfill
  \caption{\small \label{fig:tgccomb}
Anomalous nTGC 95\% confidence intervals from ATLAS, LEP~\cite{bib:LEPEW2006} and Tevatron~\cite{bib:D0_ZZ1}
experiments. Luminosities, centre-of-mass energies and cut-offs $\Lambda$ for each experiment are shown.
   }
 \end{center}
 \end{figure}

\section{Conclusions}\label{sec:Conclusions}
        A measurement of the \ZZS\ production cross section in LHC proton--proton collisions at 
$\sqrt{s}$ = 7 TeV is presented with data collected by the ATLAS detector, using the \zzSllll\ and \zzllvv\ decay channels. 
Fiducial cross sections are measured for three production and decay selections, and the results are compatible with the SM expected cross sections. 
Using the \zzllll\ and \zzllvv\ selections, the  
total \ZZ\ production cross section is determined to be: 
\begin{eqnarray*}
\sigma_{ZZ}^\mathrm{tot} &=& \textrm{\totzz}.
\end{eqnarray*}
The result is statistically consistent with the NLO Standard Model 
prediction of \theoryzzmass, calculated with \Z\ bosons with a mass between 66 
and 116 GeV, and supersedes the previous measurements made with part of the same 
dataset~\cite{ATLAS_ZZ4l:1fb2011}. 
Unfolded distributions of the fiducial cross sections are derived for the $p_{\rm{T}}^{Z}$ and $\Delta\phi(\ell^{+},\ell^{-})$ of the leading $Z$ boson 
and for $m^{ZZ}$ in the \zzllll\ selection and the $m_{\rm T}$ in the \zzllvv\ selection. 

The event yields as a function of the \pT\ of the leading $Z$ boson for the \zzllll\ and \zzllvv\ selections are used to derive 95\% confidence intervals for anomalous neutral triple gauge boson couplings. 
These limits are more stringent than those derived from measurements at 
LEP~\cite{bib:LEPEW2006} and the Tevatron~\cite{bib:D0_ZZ1}.
They improve the previous published results from ATLAS~\cite{ATLAS_ZZ4l:1fb2011} by approximately a factor of five and supersede them.

\clearpage


We thank CERN for the very successful operation of the LHC, as well as the
support staff from our institutions without whom ATLAS could not be
operated efficiently.

We acknowledge the support of ANPCyT, Argentina; YerPhI, Armenia; ARC,
Australia; BMWF and FWF, Austria; ANAS, Azerbaijan; SSTC, Belarus; CNPq and FAPESP,
Brazil; NSERC, NRC and CFI, Canada; CERN; CONICYT, Chile; CAS, MOST and NSFC,
China; COLCIENCIAS, Colombia; MSMT CR, MPO CR and VSC CR, Czech Republic;
DNRF, DNSRC and Lundbeck Foundation, Denmark; EPLANET, ERC and NSRF, European Union;
IN2P3-CNRS, CEA-DSM/IRFU, France; GNSF, Georgia; BMBF, DFG, HGF, MPG and AvH
Foundation, Germany; GSRT and NSRF, Greece; ISF, MINERVA, GIF, DIP and Benoziyo Center,
Israel; INFN, Italy; MEXT and JSPS, Japan; CNRST, Morocco; FOM and NWO,
Netherlands; BRF and RCN, Norway; MNiSW, Poland; GRICES and FCT, Portugal; MERYS
(MECTS), Romania; MES of Russia and ROSATOM, Russian Federation; JINR; MSTD,
Serbia; MSSR, Slovakia; ARRS and MVZT, Slovenia; DST/NRF, South Africa;
MICINN, Spain; SRC and Wallenberg Foundation, Sweden; SER, SNSF and Cantons of
Bern and Geneva, Switzerland; NSC, Taiwan; TAEK, Turkey; STFC, the Royal
Society and Leverhulme Trust, United Kingdom; DOE and NSF, United States of
America.

The crucial computing support from all WLCG partners is acknowledged
gratefully, in particular from CERN and the ATLAS Tier-1 facilities at
TRIUMF (Canada), NDGF (Denmark, Norway, Sweden), CC-IN2P3 (France),
KIT/GridKA (Germany), INFN-CNAF (Italy), NL-T1 (Netherlands), PIC (Spain),
ASGC (Taiwan), RAL (UK) and BNL (USA) and in the Tier-2 facilities
worldwide.


\providecommand{\href}[2]{#2}\begingroup\raggedright\endgroup

\clearpage

\begin{flushleft}
{\Large The ATLAS Collaboration}

\bigskip

G.~Aad$^{\rm 48}$,
T.~Abajyan$^{\rm 21}$,
B.~Abbott$^{\rm 111}$,
J.~Abdallah$^{\rm 12}$,
S.~Abdel~Khalek$^{\rm 115}$,
A.A.~Abdelalim$^{\rm 49}$,
O.~Abdinov$^{\rm 11}$,
R.~Aben$^{\rm 105}$,
B.~Abi$^{\rm 112}$,
M.~Abolins$^{\rm 88}$
% ...
\bigskip

% ...
$^{11}$ Institute of Physics, Azerbaijan Academy of Sciences, Baku, Azerbaijan\\
$^{12}$ Institut de F\'{i}sica d'Altes Energies and Departament de F\'{i}sica de la Universitat Aut\`{o}noma de Barcelona and ICREA, Barcelona, Spain\\
% ...
$^{21}$ Physikalisches Institut, University of Bonn, Bonn, Germany\\
% ...
$^{48}$ Fakult\"{a}t f\"{u}r Mathematik und Physik, Albert-Ludwigs-Universit\"{a}t, Freiburg, Germany\\
$^{49}$ Section de Physique, Universit\'{e} de Gen\`{e}ve, Geneva, Switzerland\\
% ...
$^{88}$ Department of Physics and Astronomy, Michigan State University, East Lansing MI, United States of America\\
% ...
$^{105}$ Nikhef National Institute for Subatomic Physics and University of Amsterdam, Amsterdam, Netherlands\\
% ...
$^{111}$ Homer L. Dodge Department of Physics and Astronomy, University of Oklahoma, Norman OK, United States of America\\
$^{112}$ Department of Physics, Oklahoma State University, Stillwater OK, United States of America\\
% ...
$^{115}$ LAL, Universit\'{e} Paris-Sud and CNRS/IN2P3, Orsay, France\\
% ...
\end{flushleft}


\end{document}